\documentclass[aps,reprint,amsmath,amssymb]{revtex4-2}
\usepackage{graphics}
\usepackage{subfigure}
\usepackage{epsfig}
\usepackage{epsf,epic}
\usepackage{dcolumn}
\usepackage{color}
\usepackage{hyperref}
\usepackage{marvosym}
\usepackage{subfigure}
\usepackage{amsmath}
\usepackage{amssymb}
\usepackage{amsfonts}
\usepackage{wrapfig}
\usepackage{pstricks}
\usepackage{multirow}
\usepackage{bm}
\graphicspath{{./figs/}}
\usepackage{dcolumn}
\newcommand{\etal}{\textit{et al.\ }}

\begin{document}
\title{Improved quasiparticle self-consistent electronic band structure and excitons in $\beta$-LiGaO$_2$}

\author{Niloufar Dadkhah}
\author{Walter R. L. Lambrecht}\email{walter.lambrecht@case.edu}

\affiliation{Department of Physics, Case Western Reserve University, 10900 Euclid Avenue, Cleveland, Ohio 44106-7079, USA}
\author{Dimitar Pashov}
\affiliation{Department of Physics, King’s College London, London WC2R 2LS, United Kingdom}
\author{Mark van Schilfgaarde}
\affiliation{National Renewable Energy Laboratory, Golden, Colorado 80401, USA}

\begin{abstract}
  The band structure of $\beta$-LiGaO$_2$ is calculated using the quasiparticle self-consistent QS$G\hat W$ method where
  the screened Coulomb interaction $\hat W$ is evaluated including electron-hole interaction ladder diagrams and $G$ is the
  one-electron Green's function. Improved convergence compared to previous calculations leads to a significantly larger band gap of about 7.0 eV.
  However, exciton binding energies are found to be large and lead to an exciton gap of about 6.0 eV if also a zero-point-motion
  correction of about $-0.4$ eV is included. These results are in excellent agreement with recent experimental results on the
  onset of absorption. Besides the excitons observed thus far, the calculations indicate the existence of
  a Rydberg-like series of exciton excited states, which is however modified from the classical Wannier exciton model by the anisotropies of the material and the more complex mixing of Bloch states in the excitons resulting from the Bethe-Salpeter equation.  The exciton fine structure and the exciton wave functions are visualized and analyzed in various ways.
 \end{abstract}

\maketitle

\section{Introduction}
Lithium gallate ($\beta$-LiGaO$_2$)  is a well-known optical material, which
has recently also received interest as a potential ultrawide-band-gap
semiconductor. Its crystal structure was reported  by Marezio \cite{Marezio65} and consists of a
cation-ordered wurtzite-derived structure with space group $Pna2_1$. It can be grown in bulk form by the
Czochralski method \cite{Ishii98,Chen14} or epitaxially on ZnO \cite{Ohkubo2002}.  It has been studied in the past for its piezoelectric
properties \cite{Nanamatsu72,Gupta76}, can be alloyed with ZnO \cite{Omata08,Omata11} and CuGaO$_2$ \cite{Suzuki2019},
and has been studied as a substrate
for GaN \cite{Ishii98,Christensen05,Doolittle98}. Its heat capacity and other thermodynamic properties were studied
by Weise and Neumann \cite{Weise96} and  Neumann \etal\cite{Neumann87}.
Various studies were also done of its phase transitions under high pressure \cite{Lei10,Lei13,Radha21}.
Its elastic, phonon, and piezoelectric properties were calculated
using density functional theory (DFT) by Boonchun and Lambrecht \cite{Boonchun10}. Its electronic structure was calculated
at the quasiparticle self-consistent (QS)$GW$ level (where $G$ is the one-electron Green's function and $W$ the screened Coulomb interaction)
\cite{Boonchun11SPIE,Radha21} and earlier at the DFT level using  the modified Becke-Johnson exchange-correlation \cite{Becke06,TranBlaha09}
functional by Johnson \etal\cite{Johnson2011}.  Its optical gap was obtained from absorption measurements \cite{Wolan98,Chen14}
and a combination of x-ray absorption and emission spectroscopies \cite{Johnson2011} and found generally to be about 5.3--5.6 eV.
Its native defects were recently studied \cite{Boonchun19}, as well as its potential for n-type and p-type doping \cite{Dabsamut20}.
It was predicted that Si and Ge would be shallow donors, while Sn would be a deep donor. Doping by various diatomic molecules was also investigated
but not found to lead to p-type doping \cite{Dabsamut22}. Electron paramagnetic resonance of Li and Ga vacancies was reported by Lenyk \etal \cite{Lenyk18} and analyzed computationally by Skachkov \etal \cite{Skachkov20}.

Recently, the infrared (phonon related) as well  as visible ultraviolet (interband transition related)
optical properties were studied with reflectivity, transmission, and spectroscopic ellipsometry by Tum\.enas \etal\cite{Tumenas17}
and indicated the existence of sharp excitons near 6.0 eV. Luminescence properties were studied by Trinkler \etal\cite{Trinkler17,Trinkler22}
and the photoluminescence excitation spectroscopy confirmed the presence of sharp free excitons near 6.0 eV.
The anisotropic splitting of these excitons, reported in \cite{Trinkler22}, reflects the valence band splitting, characteristic
of the orthorhombic symmetry of the crystal, and is in good agreement with the recent computational study
by Radha \etal \cite{Radha21}. However, the free excitons at about 6.0 eV imply a band gap significantly higher than
most previous studies indicated \cite{Wolan98,Chen14,Johnson2011}. This led us to reexamine the QS$GW$ calculations
reported in Radha \etal\cite{Radha21}. Furthermore, we here use an improved
QS$G\hat W$ method which includes vertex corrections in the polarization, calculate the dielectric functions using the Bethe-Salpeter equation approach, and study the thus obtained excitons in some detail.

\section{Computational Method} \label{sec:method}
We use here essentially the same computational method as in Radha \etal \cite{Radha21} but performed additional convergence studies and also now avoid the somewhat \textit{ad hoc} correction of the self-energy by a factor of 80 \% by using the recently developed extension of the $GW$ method in which the screened Coulomb interaction $W$  is evaluated beyond the
random phase approximation (RPA) by including ladder diagrams \cite{Cunningham18,Cunningham23,Radha-LCO21}.
The QS$GW$ method is based on the well-known many-body-perturbation theory of Hedin \cite{Hedin65,Hedin69}
but uses an iteration scheme where a nonlocal but Hermitian and energy-independent exchange-correlation potential $\tilde\Sigma$ is extracted from the $GW$ self-energy $\Sigma(\omega)$, which is used to update the noninteracting Hamiltonian $H^0$, and its Green's function $G^0$ is used to calculate the self-energy $\Sigma=iGW$ of the next iteration \cite{MvSQSGWprl}. The implementation of the method in terms of
a mixed  interstitial plane-wave product basis set and other technical aspects are detailed in Ref. \cite{Kotani07}, and
the full-potential linearized muffin-tin-orbital (FP-LMTO) band-structure method employed and integrated with the
$GW$ method is fully described in Ref. \cite{questaalpaper}, which describes the {\sc Questaal} code \cite{questaal} used in this work.
While electron-hole effects  can also be incorporated through including an exchange-correlation kernel in the inverse dielectric
function in the framework of time-dependent DFT, that approach relies on the accuracy of the kernel which
typically needs to be extracted from Bethe-Salpeter-equation (BSE) calculations \cite{Shishkin07}, or uses the bootstrap kernel \cite{ChenPasquarello15}.
The approach introduced by Cunningham \etal\cite{Cunningham18,Cunningham23} instead calculates directly the four-point generalized susceptibility
by solving a Bethe-Salpeter equation at each ${\bf q}$ point rather than only in the long-wavelength limit ${\bf q}\rightarrow{\bf 0}$.
It does so only for $W(\omega=0)$ and within the Tamm-Dankoff approximation (TDA) but then contracts the four-point generalized susceptibility back to  the
two-point polarizability, $P(12)=P_{RPA}(12)-\int P_{RPA}(1134)W(34,\omega=0)P(3422) d(34)$,  needed to evaluate $W=(1-Pv)^{-1}v$.
The thus obtained improved screened Coulomb interaction is here denoted by $\hat W({\bf q},\omega)$. Details of the approach can also be found in \cite{Radha-LCO21} where it was applied to the case of LiCoO$_2$. The approach was shown to be equivalent \cite{Starke2012,Maggio2017} to including a vertex correction to the polarizability propagator  extracted from the functional derivative $\delta\Sigma^{GW}/\delta G$
with $\Sigma^{GW}$ the $GW$ self-energy within the general Hedin set of equations. However, we clarify that no vertex corrections are included in the self-energy itself,
which is justified in part by cancellations of the
$Z$ factor in $G=ZG^0$+$\tilde G$~\cite{Kotani07}, which measures the quasiparticle versus the incoherent part ($\tilde G$) of the one-particle Green's function,
and in the vertex which behaves as $\Gamma\rightarrow 1/Z$ in the low-frequency, $\omega{\rightarrow}0,\ q{\rightarrow}0$ limit.
This cancellation applies whenever the noninteracting $G^0$ is used as opposed to the fully self-consistent $G$.
QS$GW$ and QS$G\hat{W}$ both make use of it, and as shown in Ref.~\cite{Cunningham23} it does a remarkably good job at predicting both the band gap and $\varepsilon_\infty$ for a wide range of materials systems.
After calculating the band structure in the GGA using the PBEsol functional \cite{PBEsol}
as a starting point, QS$GW$ (which becomes independent of the starting point)
with $W$ calculated in RPA, and QS$G\hat W$ with $\hat W$ calculated including the ladder diagrams as detailed above, we calculate the optical dielectric function following closely the usual BSE approach \cite{Onida02}. Specifically, we use the modified response function \cite{Hanke78}
\begin{equation}
  \bar{P}(1234)=P^0(1234)+\int d(5678)P^0(1256)K(5678)\bar{P}(7834) \label{eq:BSE1}
\end{equation}
with the  kernel
\begin{equation}
  K(1234)=\delta(12)(34)\bar{v}-\delta(13)\delta(24)\hat{W}(12). \label{eq:kernel}
\end{equation}
with 
$\bar{v}_{{\bf G}}({\bf q})=4\pi/|{\bf q}+{\bf G}|^2$ if ${\bf G}\ne0$
and zero otherwise. 
The macroscopic dielectric function is then given by
\begin{equation}
  \varepsilon_M(\omega)=1-\lim_{{\bf q}\rightarrow0}v_{{\bf G}=0}({\bf q})\bar{P}_{{\bf G}={\bf G}^\prime=0}({\bf q},\omega) \label{eqepsmac}
\end{equation}
Note, that unlike the usual approach, we here use $\hat W$ in Eq. \ref{eq:kernel} rather than the RPA $W$.

\section{Results}
\subsection{Energy bands}
\begin{figure}
  \includegraphics[width=8cm]{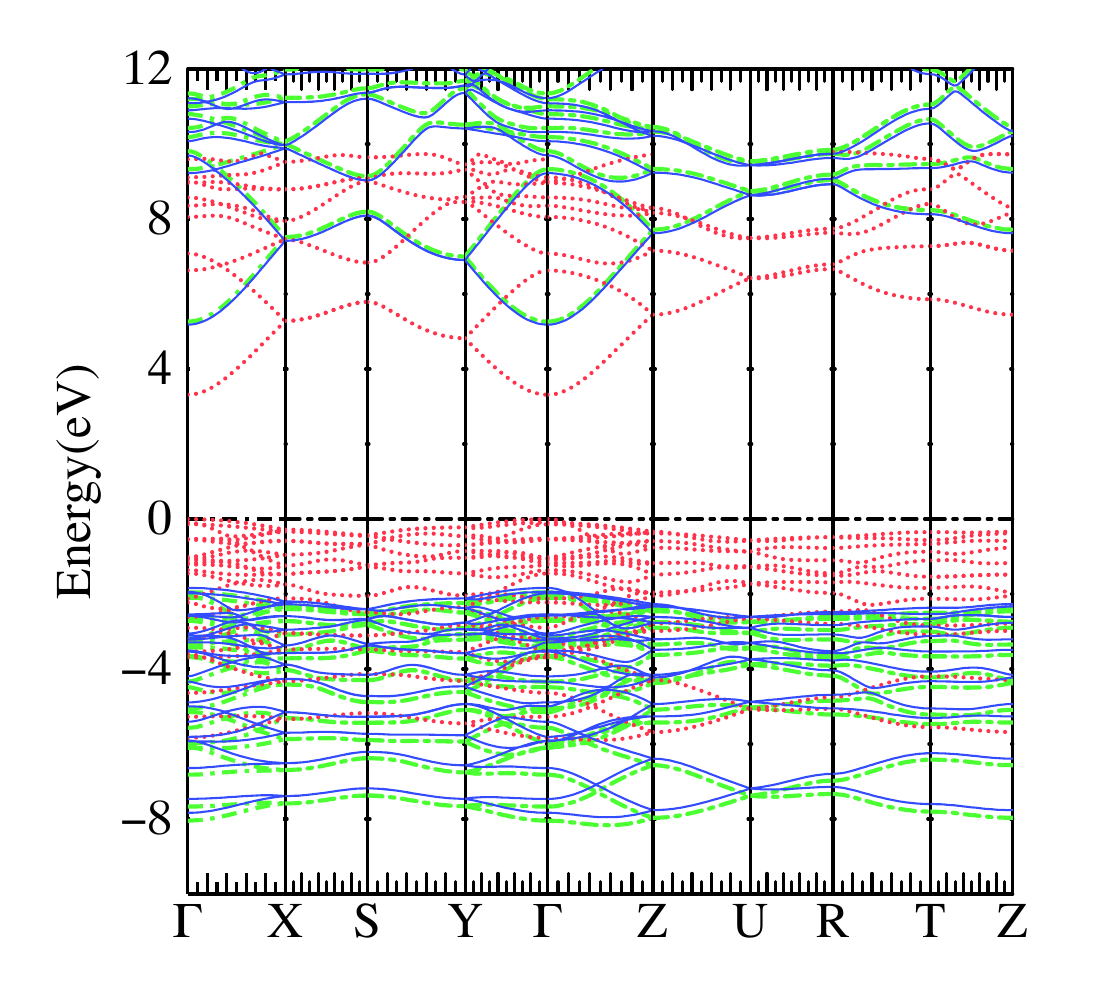}
  \caption{Band structure of LiGaO$_2$ in the GGA (PBEsol) (red dots), QS$GW$ (green dashed), and QS$G\hat W$. The bands are all referred to the
    valence band maximum of the GGA band structure, thereby showing how much the gap change occurs in the valence and conduction bands separately.\label{figbnds}}
\end{figure}

\begin{table}
  \caption{Band gap of LiGaO$_2$  in different methods.}
  \begin{ruledtabular}
    \begin{tabular}{ld}
      Method & \multicolumn{1}{c}{$E_g$ (eV)} \\ \hline
      PBEsol & 3.31 \\
      QS$GW$   & 7.22 \\
      QS$G\hat W$ & 7.02 \\
      QS$G\hat{W}+$ZPR & 6.66 \\ \hline
      PBE\footnote{From Rahda \etal \cite{Radha21}} & 3.36 \\
      QS$GW^a$                                      & 6.46 \\
      0.8$\Sigma$ QS$GW^a$                            & 5.81 \\
      $GW_0$ \footnote{From Fang \cite{Dangqi22}} & 5.995 \\
      $GW_0+$ZPR$^b$ & 5.633 \\

    \end{tabular}
  \end{ruledtabular}\label{tabgap}
\end{table}

In Fig. \ref{figbnds} we show the band structure calculated at the experimental lattice parameters in the three approaches just mentioned.
We can see that using $\hat W$ only slightly changes the gap and mostly by shifting the valence band maximum (VBM) slightly back up compared to the down shift occurring in
QS$GW$ using the RPA $W$ compared to GGA.
The gaps are summarized in Table \ref{tabgap}.
Our results here differ from Ref. \cite{Radha21} even for the QS$GW$ case. We found that was due to converging the root mean square deviation of the self-energy from one iteration to the next  only to a tolerance of 10$^{-3}$ in that paper
whereas now it is converged to 10$^{-6}$. Apparently this still affects the band gap to the order of 0.1 eV. We further tested the convergence by using a
$4\times4\times4$ mesh instead of $3\times3\times3$ for the calculation of the self-energy but this was found to change the QS$GW$ gap from 7.218 to 7.207 eV, so
the $3\times3\times3$ mesh was deemed converged to $\pm0.01$ eV and used for the subsequent calculation of $\hat W$. The reduction of the self-energy
shift owing to the ladder diagrams can be taken as $(E_g(QSG\hat W)-E_g(GGA))/(E_g(QSGW)-E_g(GGA))$ and amounts to 0.948, so a reduction by only $\sim$5 \%.

The zero-point motion band gap renormalization (ZPR) due to electron-phonon coupling  also needs to be considered. This correction  is
dominated by the longitudinal optical phonon Fr\"ohlich interaction and was  estimated in Radha \etal\cite{Radha21}
to be about $-0.2$ eV. It was recently calculated explicitly by Fang \cite{Dangqi22} to be $-0.36$ eV including all phonons  and $-0.31$ eV using only the Fr\"ohlich contribution.
This author also performed $GW_0$ calculations and obtained a gap of 5.995 eV without  and 5.633 eV with ZPR.
Adding the ZPR correction to our larger gap, the quasiparticle gap is here obtained to be 6.66 eV.

\begin{figure}
  \includegraphics[width=9cm]{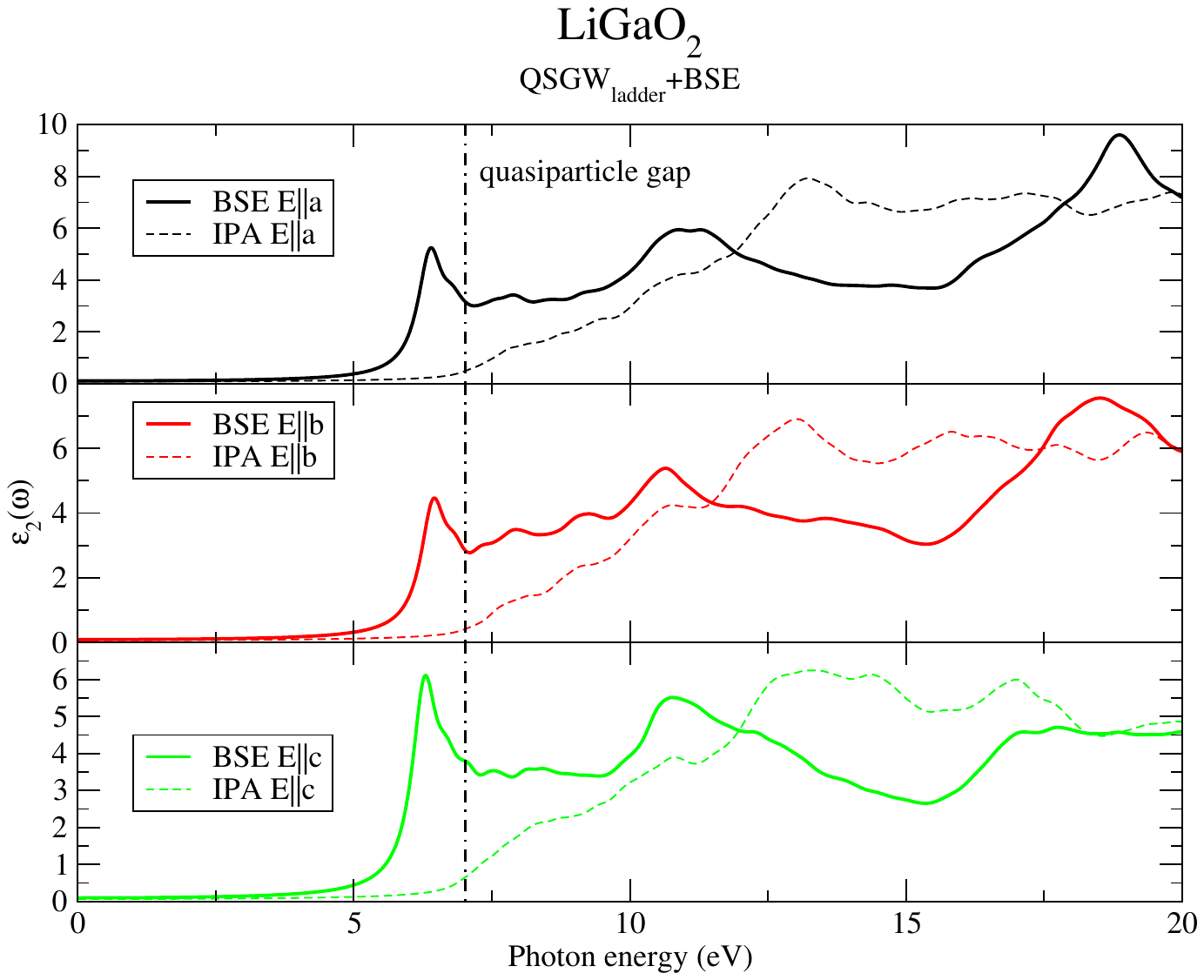}
  \caption{Imaginary part of the macroscopic dielectric function tensor  $\varepsilon_2(\omega)_{\alpha\alpha}$ for three polarizations $\alpha$ within
    the BSE and independent-particle approximation (IPA).  The quasiparticle gap is indicated by the dashed line.  \label{figeps}}
\end{figure}

\begin{figure}
  \includegraphics[width=9cm]{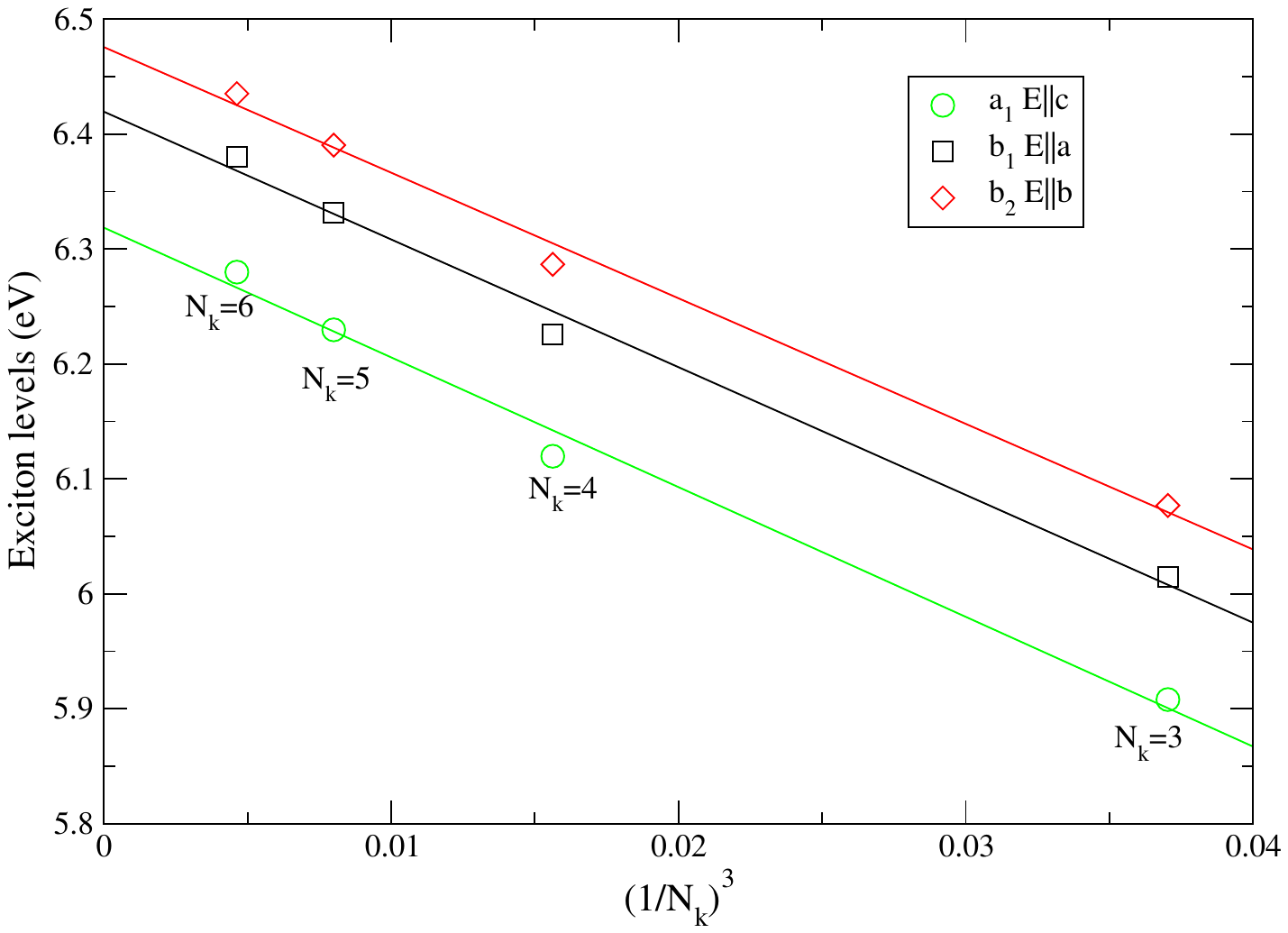}
  \caption{Exciton band gaps for each polarization as function of {\bf k}-mesh density. \label{figexciton}}
\end{figure}
\subsection{Dielectric function and excitons}
Next we calculate the macroscopic dielectric function using the BSE method. $\hat W$, as defined in Sec. \ref{sec:method}, is used in  Eq. (\ref{eq:kernel}).
The results are shown in Fig. \ref{figeps}. One can see that compared to the independent-particle approximation (where neither electron-hole nor local field effects are included)
the shape of the dielectric function is strongly affected with peaks in the continuum being redshifted and a sharp exciton peak occurs below the  gap for each polarization
direction.

However, to extract an accurate exciton binding energy, it is important to converge the {\bf k}-point mesh used in the BSE two-particle Hamiltonian
$H_{vc{\bf k},v^\prime c^\prime{\bf k}^\prime}$ (see \cite{Cunningham23}).
The results for the lowest bright exciton of each polarization as function of the inverse of the number of {\bf k} points in the Brillouin zone  are shown in Fig. \ref{figexciton}.
We here used $N_k\times N_k\times N_k$ meshes with $N_k\in\{3,4,5,6\}$.  We can see that the $6\times6\times6$ mesh is close to being converged and the linear extrapolation yields the values
given in Table \ref{tabexciton}. The line before the last line in this table gives the final exciton gaps after subtracting the ZPR correction of the gap. These values agree well with
the experimental values of Trinkler \etal\cite{Trinkler22}.  Of course, there remains some uncertainty in our calculations
resulting from the extrapolations and various other approximations, such as completeness of basis set. We estimate
these to be of order 0.1 eV.

The polarization dependence results from the splitting of the valence band maximum with the $a_1$ state (corresponding to $z$ along ${\bf c}$)
forming the VBM, followed by the $b_1$ state (polarized along $x$ or ${\bf a}$) and $b_2$ (polarized along $y$ or ${\bf b}$).
Our calculated splittings for these excitons are 100 meV for the $a_1-b_1$ splitting and 155 meV for the $a_1-b_2$ splitting, whereas the corresponding
band splittings are 106 meV and 147 meV and the experimental splittings are 102 meV and 136 meV.  The closeness of the band splittings from the exciton
eigenvalues indicates that the exciton binding energy is almost constant and $\sim$0.70 eV. This is a remarkably high value. In the present calculation
only electronic screening is included in the exciton binding energy. While the gaps themselves were shifted by a ZPR, the $\hat W$
only includes electronic screening without a contribution from the lattice polarization.  However, this is justified by the final binding energies being
much larger than the highest phonon energies. The phonons thus are too slow to contribute to the screening of the electron-hole correlated motion in the bound exciton.

\begin{table}
  \caption{Exciton gap convergence and comparison with experiment. \label{tabexciton}}
  \begin{ruledtabular}
    \begin{tabular}{lddd}
      $N_k$ & \multicolumn{1}{c}{${\bf E}\parallel{\bf c}$} &  \multicolumn{1}{c}{${\bf E}\parallel{\bf a}$} &  \multicolumn{1}{c}{${\bf E}\parallel{\bf b}$} \\ \hline
      3     & 5.908 & 6.015 & 6.077 \\
      4     & 6.120 & 6.226 & 6.287 \\
      5     & 6.230 & 6.332 & 6.390 \\
      6     & 6.280 & 6.380 & 6.435 \\ \hline
      $\infty$ & 6.32 & 6.42 & 6.48 \\
      $\infty$+ZPR & 5.96 & 6.06 & 6.12 \\
      Expt.\footnote{From Trinkler \etal\cite{Trinkler22}} & 5.931 & 6.033 & 6.067
          \end{tabular}
  \end{ruledtabular}
\end{table}
\begin{figure}
  \includegraphics[width=9cm]{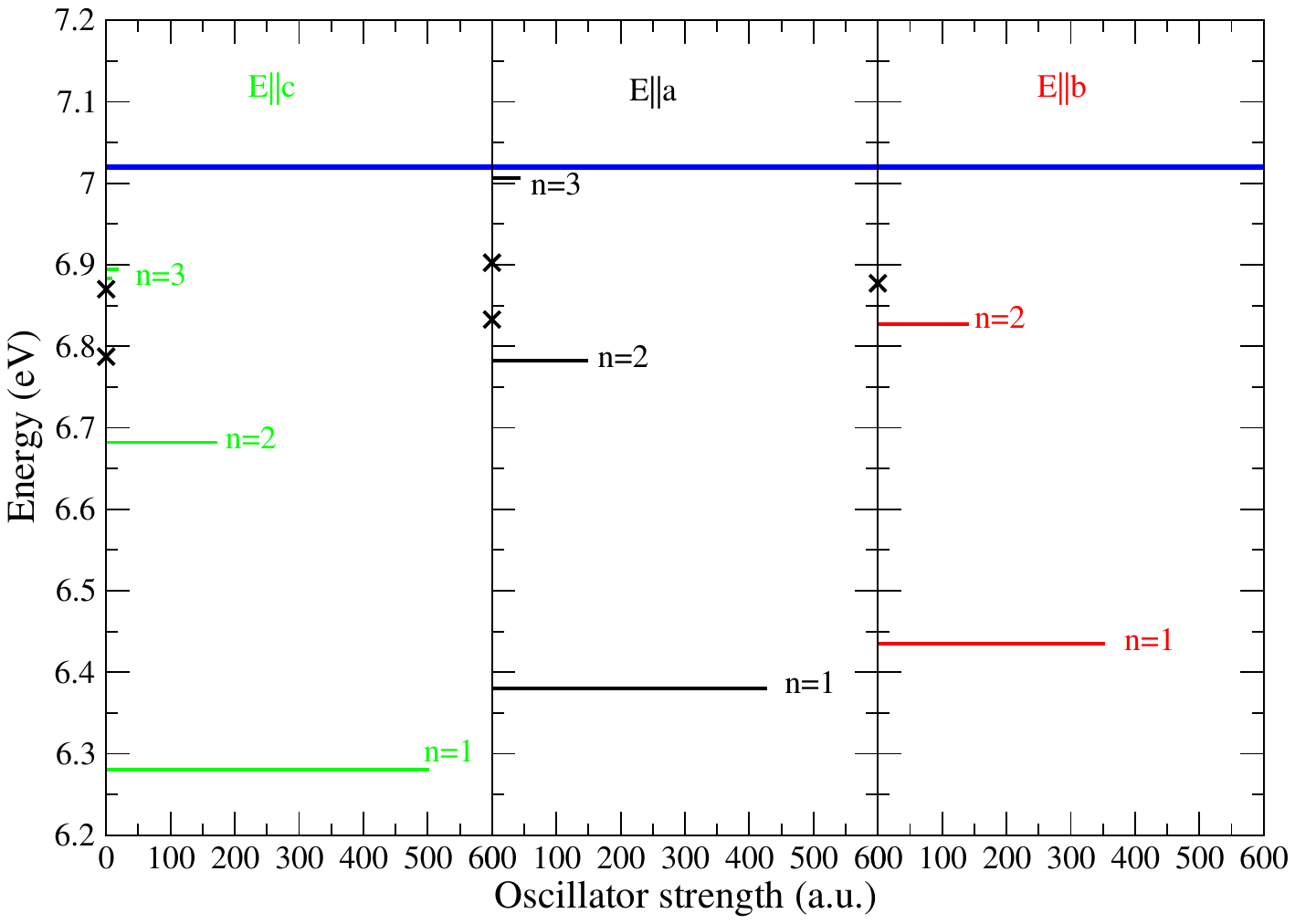}
  \caption{Exciton levels with their oscillator strengths
    for the three directions. Dark excitons indicated by crosses, placed
    arbitrarily in one of the three panels. The wide blue line indicates
    the QS$G\hat W$ gap. \label{figexlevel}}
\end{figure}

\begin{table}
  \caption{Exciton eigenvalues $\epsilon_i$ and their polarization $\lambda_i$, oscillator strength $f_i$ (arbitrary units) and
    Rydberg series quantum number $n$  \label{tabexseries}}
  \begin{ruledtabular}
    \begin{tabular}{cccc}
      $\epsilon_i$ (eV) & $\lambda_i$ & $f_i$   & $n$\\ \hline
      6.2800 & c & 502 & 1 \\
      6.3804 & a & 428 & 1  \\
      6.4352 & b & 353 & 1 \\
      6.6822 & c & 173 & 2 \\
      6.7824 & a & 149 & 2 \\
      6.7873 &   & dark & 2  \\
      6.8269 & b & 141  & 2 \\
      6.8328 &   & dark  & 2\\
      6.8700 &   & dark  & 2\\
      6.8773 &   & dark  & 2\\
      6.8836 & c & 10 & 3 \\
      6.8944 & c & 20 & 3 \\
      6.9026 & & dark & 3 \\
      7.0066 & a & 45 & 3 \\
    \end{tabular}
  \end{ruledtabular}
\end{table}

\begin{figure*}
  (a)\includegraphics[width=0.45\textwidth]{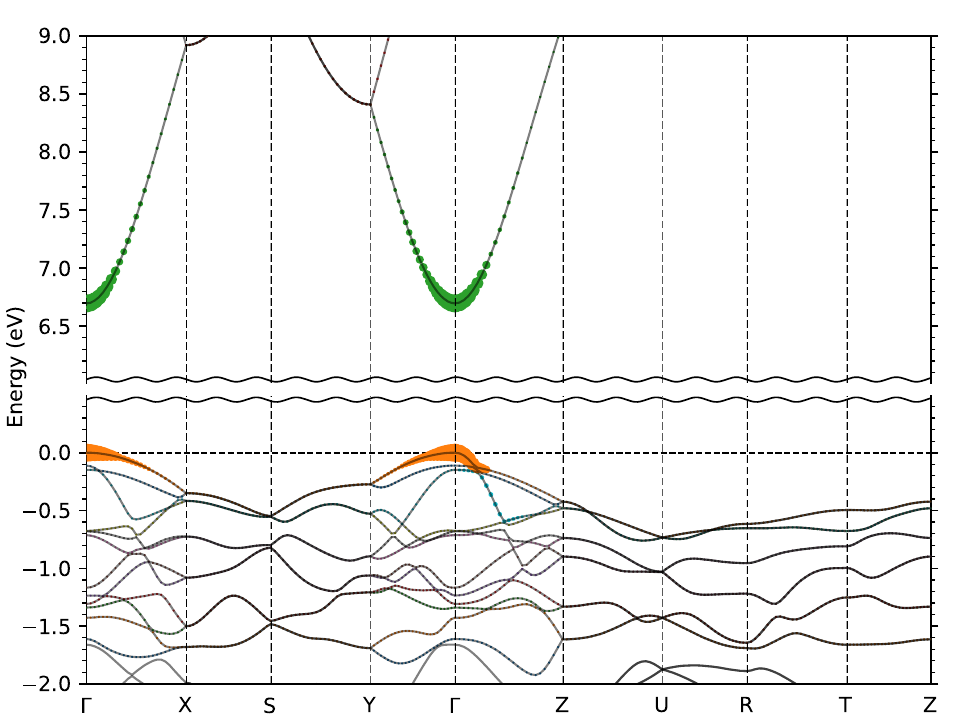}
  (b)\includegraphics[width=0.45\textwidth]{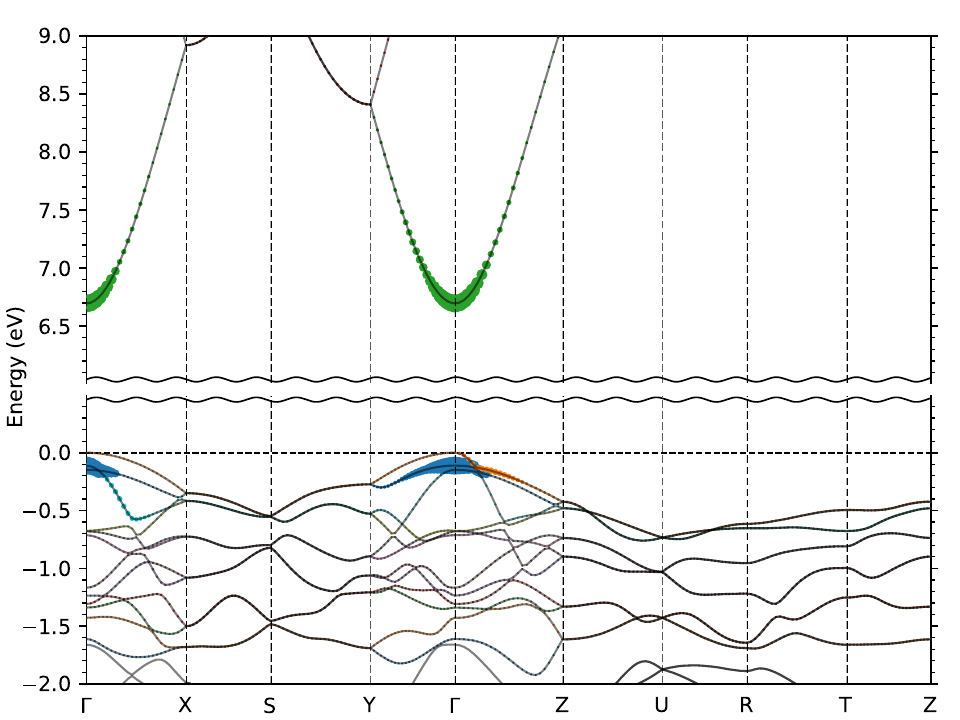}
  (c)\includegraphics[width=0.45\textwidth]{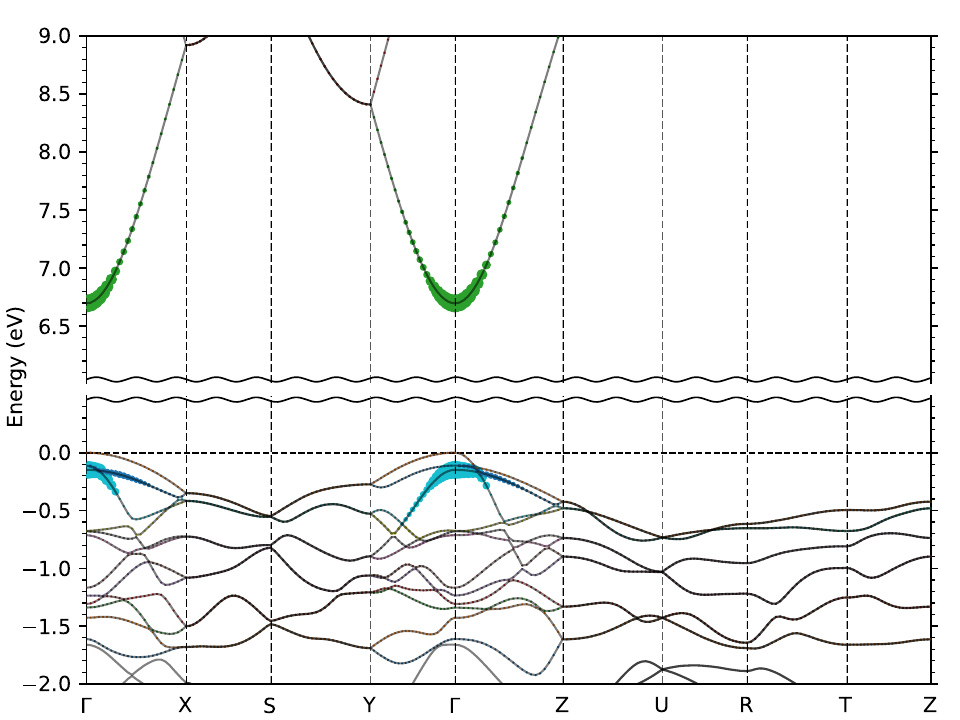}
  (d)\includegraphics[width=0.45\textwidth]{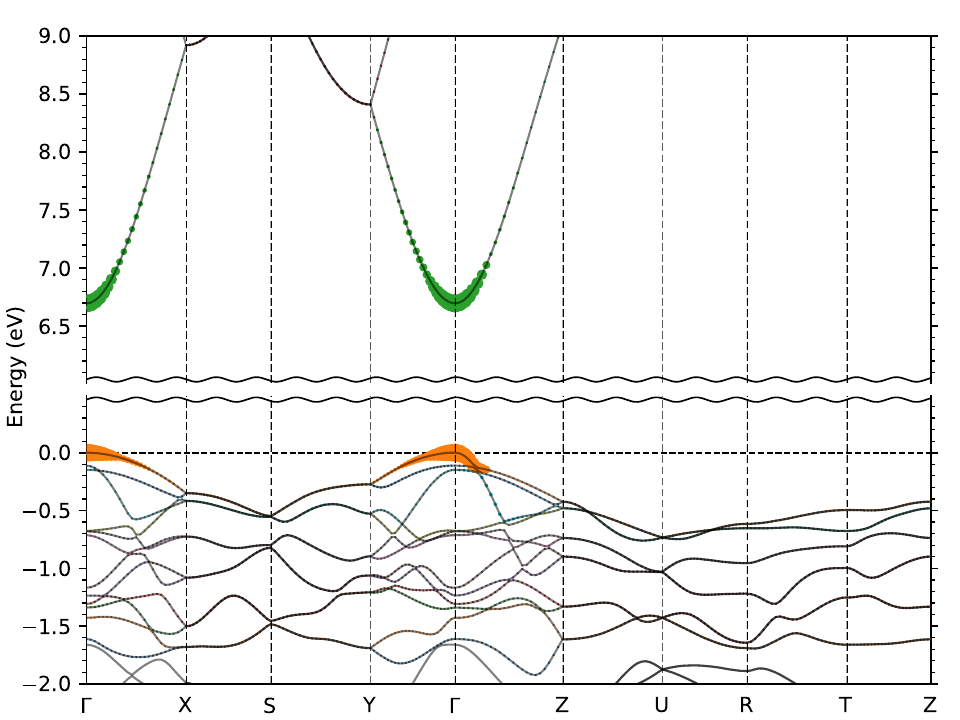}
  \caption{Weights of exciton wave function contributed by different bands
    for (a)--(d) first to fourth excitons. The size of the colored circles indicates the exciton weight $|A^\lambda_{vc{\bf k}}|^2$ for a given exciton $\lambda$. The colors 
    have no meaning and only serve to distinguish different bands. 
    \label{figex1k}}
\end{figure*}

\subsection{Exciton series analysis}
Besides the lowest energy excitonic peaks  discussed until now, we find a series of excited exciton states below the fundamental gap.
An overview of these exciton energies and their polarization is
given in Table \ref{tabexseries}. From the analysis of these excitons
given below, it becomes clear that these represent a modified Rydberg series.

First, their oscillator strengths show that a series of excitons with a
well-defined predominant polarization exist and are associated with the
top three valence band holes forming an exciton all with the same
conduction band minimum (CBM) at $\Gamma$. These are shown in
Fig. \ref{figexlevel}. They show a decreasing oscillator strength
as we move up in the series closer to the gap.  While not exactly
corresponding to the hydrogenic Rydberg series where the binding
energies would fall off as $1/n^2$, they approximately follow
a similar series. The exciton binding energy of the ground state excitons
is about 0.7 eV independent of polarization (or corresponding VB).
The difference in energy to their first excited state
is about 0.4 eV rather than 3/4 of 0.7 eV which would amount to 0.525 eV.
For the second excited state of each polarization it is about 0.6 eV,
which is rather close to 8/9 of the ground-state binding energy, which would
amount to 0.62 eV. For a hydrogenic series one would expect the oscillator
strengths to fall off as $1/n^3$. Here the oscillator strengths seem to
fall off somewhat slower with $n$. We hence tentatively label these exciton series
by a quantum number $n$ corresponding to their envelope function.
Besides these bright excitons with well-defined polarization, we also
find several dark excitons.

There are several reasons why the hydrogenic model is not expected to
apply strictly. First, the long-range screened Coulomb interaction
in an anisotropic (orthorhombic) medium is given by
\begin{equation}
  W({\bf r})=\frac{1}{\sqrt{\varepsilon_{xx}\varepsilon_{yy}\varepsilon_{zz}}}\frac{e^2}{\sqrt{\frac{x^2}{\varepsilon_{xx}}+\frac{y^2}{\varepsilon_{yy}}+\frac{z^2}{\varepsilon_{zz}}}}
\end{equation}
or in tensor notation  $W({\bf r})=e^2/\sqrt{\mathrm{det}(\varepsilon) \varepsilon_{ij}^{-1}x_ix_j}$. However, the anisotropy of the dielectric constant
$\varepsilon_\infty$ is rather small, as shown by the experimental values
extracted from the extrapolation of the
index of refraction in the range 1-4 eV to zero frequency but not including
the phonon contributions. They are
$\varepsilon_{xx}=3.027$, $\varepsilon_{yy}=2.931$, and $\varepsilon_{zz}=3.017$
\cite{Tumenas17}.

  The BSE calculations presented here also give us the real part of the
  electronic contribution to the macroscopic dielectric tensor $\varepsilon({\bf q}\rightarrow0,\omega=0)$. These values are more sensitive to the accuracy of
  the optical matrix elements than the peak positions in the spectrum, which suffer from  the difficulties in evaluating the contributions of the nonlocal
  self-energy $d\Sigma/dk$. The latter represents an additional term to the momentum operator ${\bf p}/m$ in the
  commutator  $[{\bf r},H]$ giving the velocity operator.
  To bypass this problem they are calculated
  at finite ${\bf q}$ and extrapolated to ${\bf q}\rightarrow0$ using
  a model dielectric function \cite{Cappellini93} for the
  $q$ dependence of the form
  $[\varepsilon(q)-1]^{-1}=[\varepsilon(0)-1]^{-1}+\alpha q^2+\beta q^4$. This procedure using a few {\bf q} points near ${\bf q}=0$ in each direction gives
  $\varepsilon_{xx}=2.90$, $\varepsilon_{yy}=2.83$, $\varepsilon_{zz}=2.88$,
  using 24 valence bands and 12 conduction bands. The results depend
  slightly on how we interpolate. Using a quadratic interpolation on $\varepsilon(q)$ directly gives $\varepsilon_{xx}=2.99$, $\varepsilon_{yy}=2.95$ $\varepsilon_{zz}=2.96$
  even closer to the experimental results. No matter which extrapolation to
  $q=0$ is used, 
  these are robustly within $\sim 2$\% of the experimental values, similar to the findings of
  Ref.~\cite{Cunningham23} for a wide range of materials systems. 
  In contrast, if we use the $W$ without ladder diagrams
    in the BSE, and start from the QS$GW$ self-energy, we find
    $\varepsilon_{xx}=2.81$,  $\varepsilon_{yy}=2.75$, $\varepsilon_{zz}=2.81$,
    which are systematically smaller by $\sim3$\%  than using $\hat W$ indicating the underscreening of $W$ in the
    standard QS$GW$, and which in the present material is consistent with the corresponding overestimate of the $\Sigma$ or gap correction
    by about 5 \% when using $W$
    instead of $\hat W$.  In several materials, this error in the screening is somewhat larger, of order 10-20 \%.  But the point is that the $\varepsilon(q\rightarrow0,\omega=0)$ is underestimated in the same systematic way as the $\Sigma$
    is overestimated. 
  This provides another
  strong indication that the quasiparticle gaps obtained in the
  ladder approximation (QS$G\hat W$) include the right amount of screening.
  Both the peak positions of the excitons and the real parts $\varepsilon_1(0)$
  agree well with experiment.

  Returning to the discussion of the applicability of the hydrogenic model for
  the excitons, we note that
the effective mass tensor is strongly anisotropic for each VB with a small
effective mass of order 0.4 for the direction corresponding to the symmetry
of the state, and mass of order 3.5-3.8 in the other directions.
For example, for the VBM of symmetry $a_1$ corresponding to $z$, the
mass is small in the $z$ direction but large in the $x,y$ directions. Likewise
for the next two valence bands. The conduction band mass is nearly isotropic
and close to 0.4 eV. So, the reduced mass is about 0.2 for the direction
with the small valence band mass and about 0.4 for the other directions.
The kinetic energy in the relative motion equation of electron and hole
would be $\frac{\hbar^2}{2}\frac{1}{\mu_i}\frac{\partial^2}{\partial x_i^2}$, with $\mu_i$ the reduced mass component $i=x,y,z$ and with summation convention.
We thus expect excitons of a given symmetry to be somewhat more extended in
the direction of the small reduced mass.
Next, for excitons closer
and closer to the gap, or with smaller binding energies, {\bf k}-point
convergence becomes more and more challenging and requires a finer mesh.
So, there are increasing errors due to the {\bf k}-mesh coarseness
for higher excited state excitons. Finally, the excitons are strictly not
corresponding to a single {\bf k} point and symmetry but are
a mixture of states of different {\bf k}.

\begin{figure}
  (a)\includegraphics[width=0.45\textwidth]{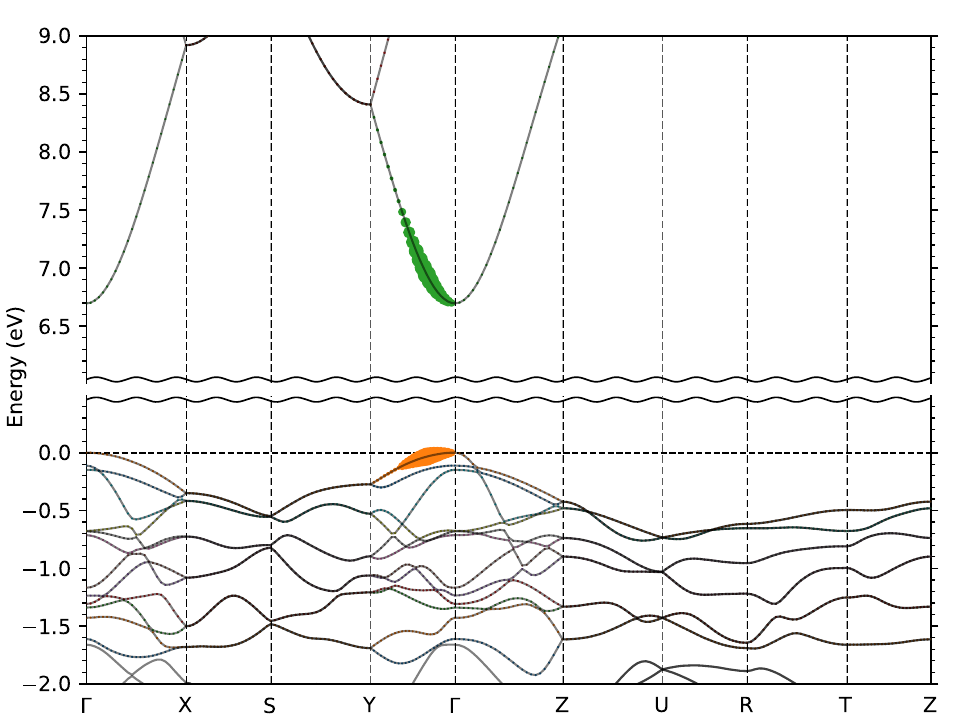}
  (b)\includegraphics[width=0.45\textwidth]{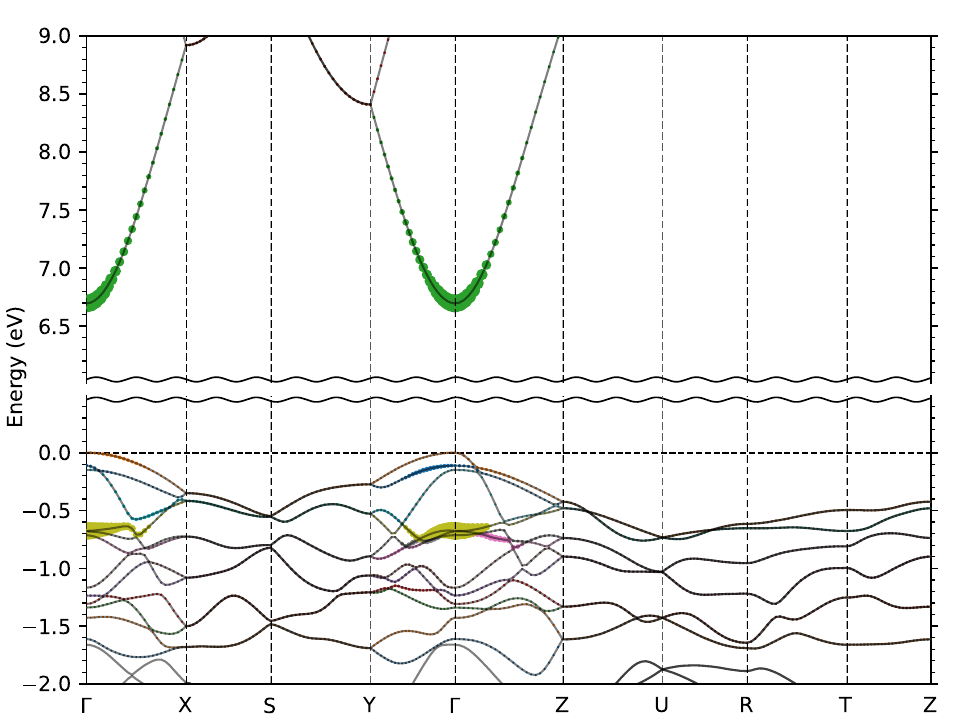}
  \caption{Band weights of the lowest (a)  and highest (b)
    energy dark excitons.\label{figdark1}}
\end{figure}

\subsection{Exciton visualization}
To verify the  association of the excitons with a Rydberg-like series
of different envelope functions and to better understand the dark
excitons, we use three different approaches.
First, we analyze the excitons  by considering which band-to-band
transitions primarily contribute to each exciton and how these are distributed
in {\bf k} space by showing their intensity on the band structure plot.
This is shown for the first four excitons in Fig. \ref{figex1k}.
In spite of the relatively high exciton
binding energy, these excitons are clearly Wannier-like with the main
contributions coming from the CBM near $\Gamma$ and for excitons
1, 2, 3 the corresponding valence bands 1, 2, 3 counted
from the VBM downward.  The zoom-in near the VBM for the first, second, and third
excitons confirms that they are coming from the
top three valence bands with $a_1$, $b_1$, $b_2$  symmetry, respectively.
One thus expects these excitons  to be delocalized in real space.
Similarly, we also
find the fourth exciton to arise from the top valence band, which
clearly identifies it as part of the Rydberg series of excitons
related to this band edge. Further analysis of the {\bf k}-space
distribution is given later and shows that it has a {\bf k}-space
envelope function
with a radial node whereas the first exciton has a nodeless envelope function
but in Fig. \ref{figex1k} this is not visible and exciton 1 and
exciton 4 appear identical.

Now, looking at the dark excitons, Fig. \ref{figdark1} shows that the first
dark exciton (at 6.7873 eV)
has zero contribution from $\Gamma$ and has contributions only
from the top valence and lowest conduction bands but only along
the $\Gamma-Y$ symmetry
line. This is readily explained if it is a 2$p_y$-like envelope function
which has a nodal plane  in the $x$ plane. While spherical symmetry
does not strictly apply, as already discussed above, we can still
classify the excitons according to the irreducible representations
of the point group at $\Gamma$ in so far as the excitons are
dominated by contributions from band-to-band pairs at $\Gamma$.
Thus a $p_y$ spherical symmetry corresponds to $b_2$ symmetry in the $C_{2v}$ group and is characterized by odd symmetry relative to the $xz$ mirror plane
perpendicular to $y$.

Similarly (not shown), the second dark exciton (at 6.8328 eV) also has
contributions from the top valence
band and bottom conduction band but now has a node in the $y$ plane,
so it must have a envelope function with approximately
$p_x$ spherical harmonic character,
or more precisely, $b_1$ symmetry in the $C_{2v}$ point group.
The next dark exciton (at 6.8700 eV) has contributions from the second
valence band (which has $b_1$ or $x$-like symmetry) but has contributions
only along $\Gamma-Y$ so it has an $x$-nodal plane.
Finally, the highest energy dark exciton (closest to the conduction band)
is shown in Fig. \ref{figdark1} because it has a somewhat different
interesting character. One can see that here several valence bands participate.
The second valence band has contributions along both $\Gamma-X$
and $\Gamma-Y$ but not at $\Gamma$ and not along $\Gamma-Z$.
We thus conclude it has a $z$-nodal plane  in terms of
these contributions. However, it also has contributions from a deeper valence
band which has contributions in all three directions near $\Gamma$
but is nonetheless dark because of the symmetry of this band, which
we checked to be $a_2$ and therefore not dipole allowed.

\begin{figure*}
  (a)\includegraphics[width=3cm]{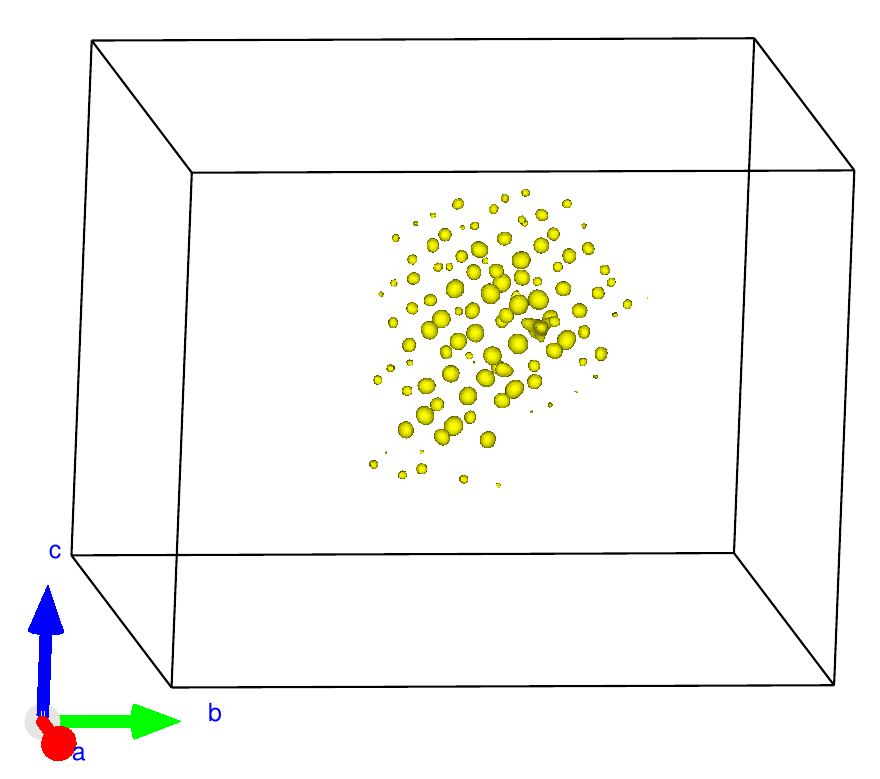}(b)\includegraphics[width=3cm]{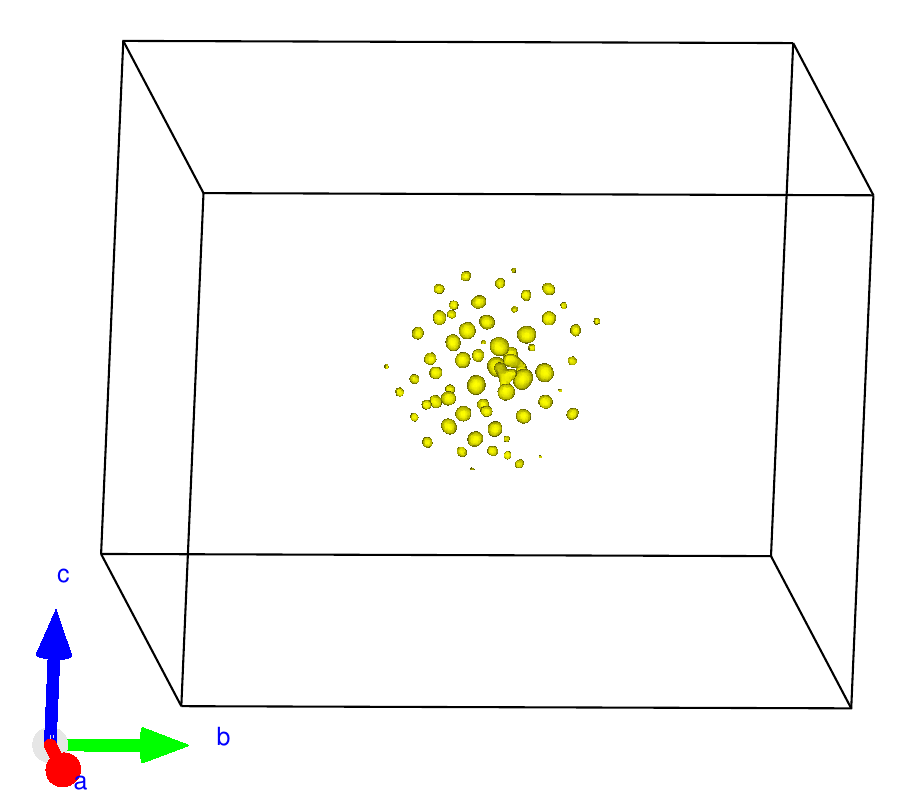}(c)\includegraphics[width=3cm]{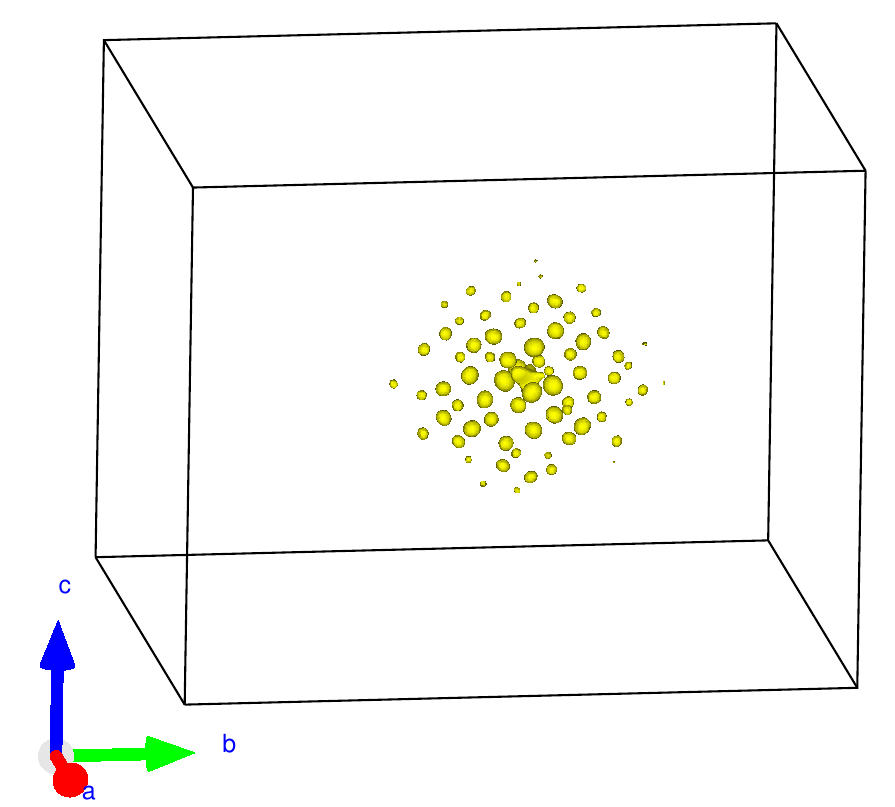}
  (d)\includegraphics[width=3cm]{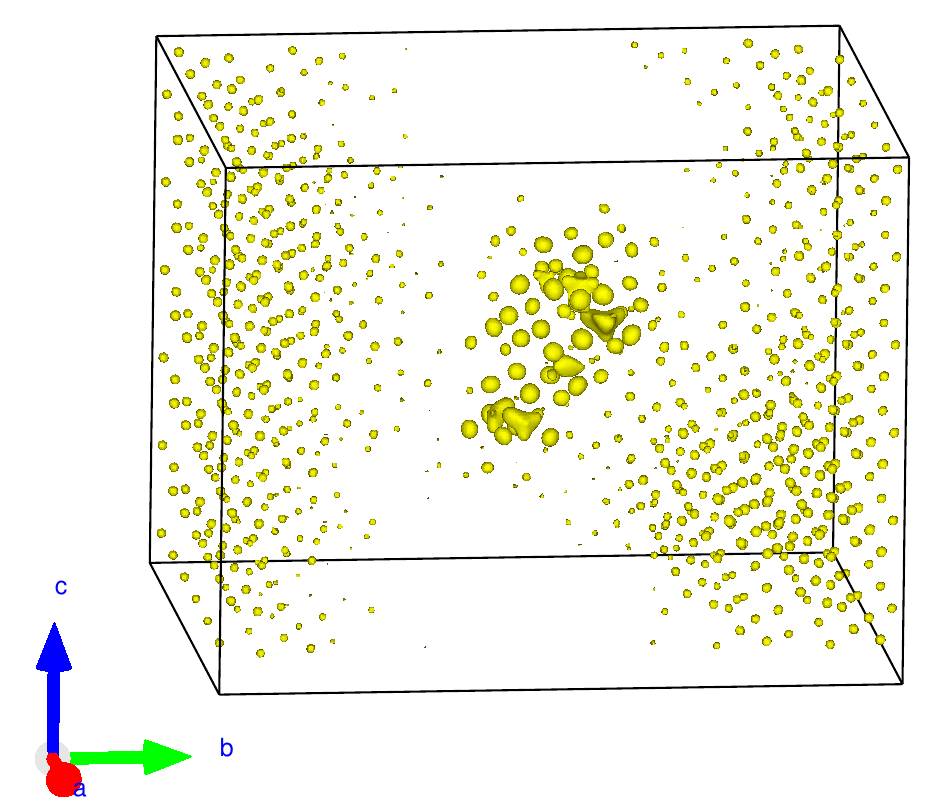} (e) )\includegraphics[width=3cm]{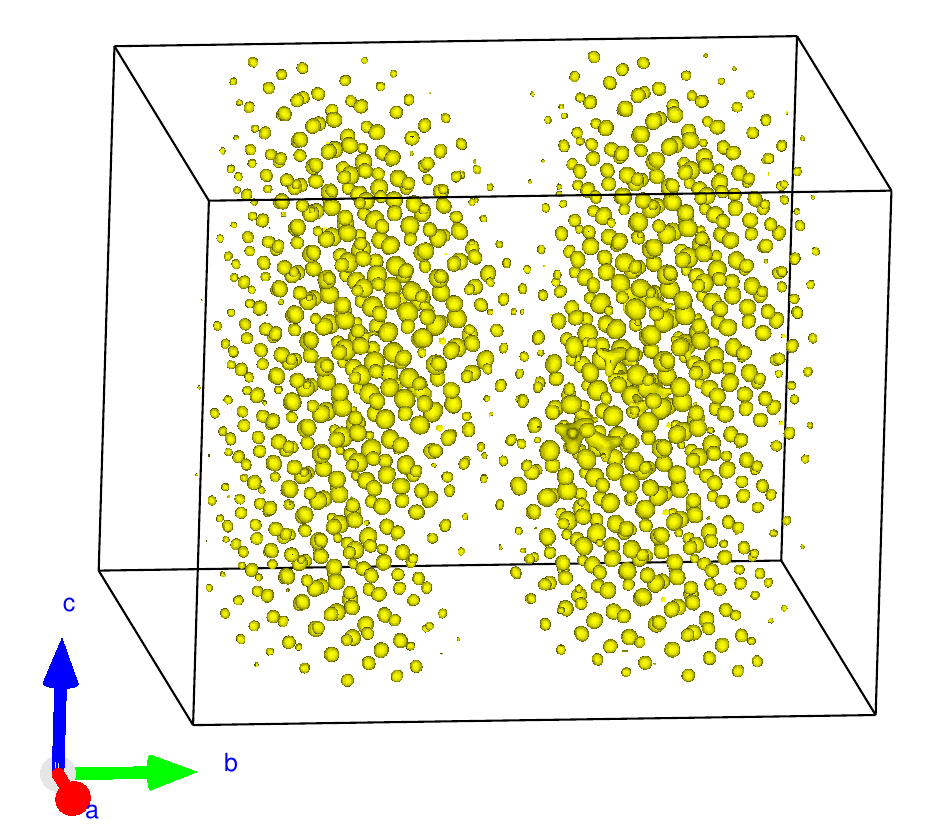}
  \caption{Spatial distribution of excitons: the yellow isosurfaces
    correspond approximately to 10 \% of the maximum value. (a)--(d) correspond to excitons 1--4 and (e) to the exciton No. 6 in Table
    \ref{tabexseries}, which is the first dark exciton.\label{figrealex}}
\end{figure*}

\begin{figure*}
  
  (a1)\includegraphics[width=4cm]{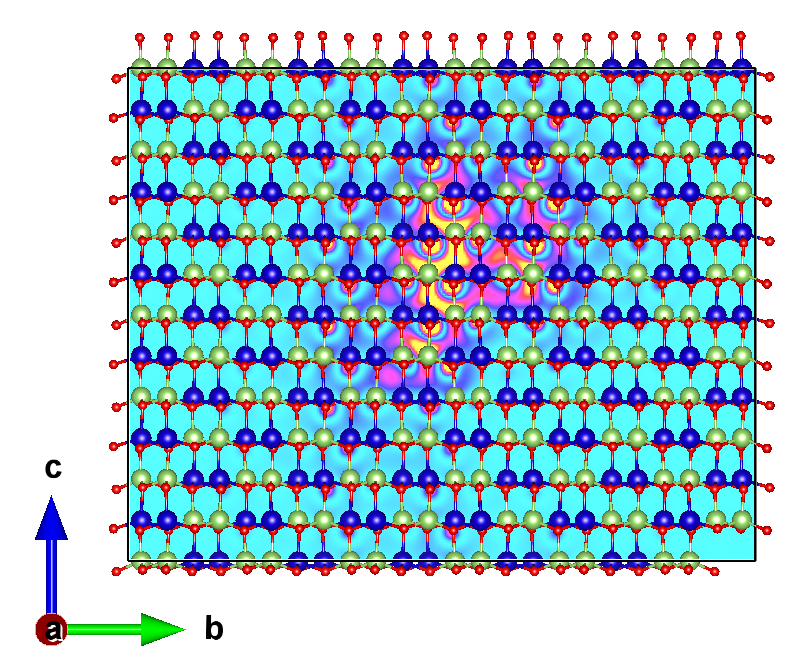}(a2)\includegraphics[width=4cm]{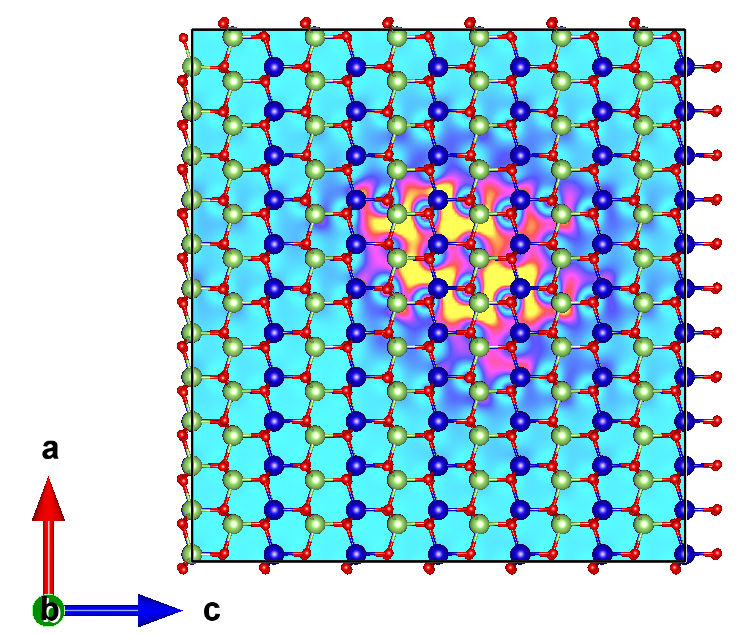}(a3)\includegraphics[width=4cm]{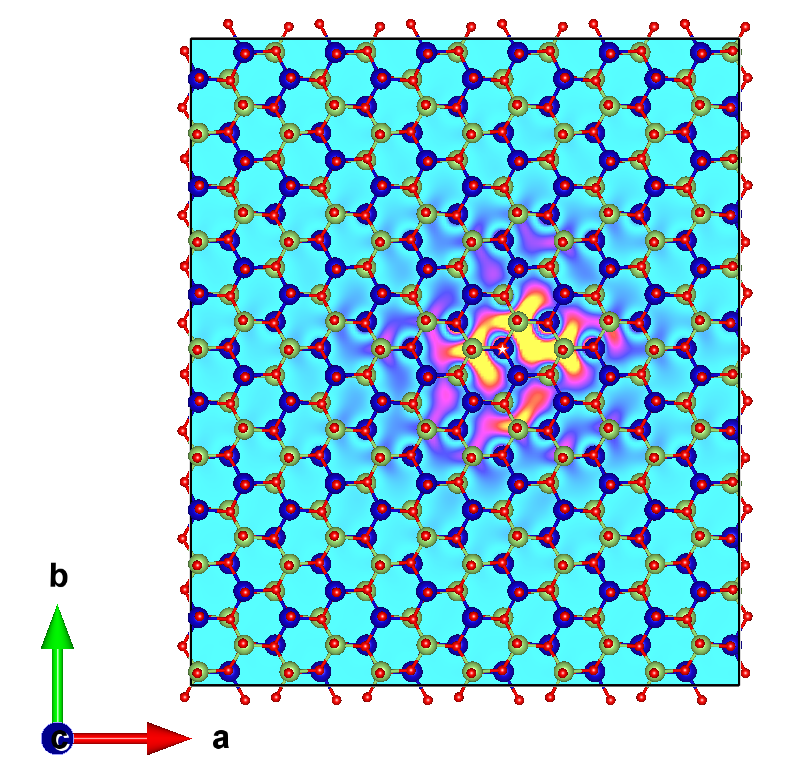}\includegraphics[height=5cm]{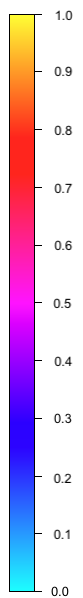}
  (b1)\includegraphics[width=4cm]{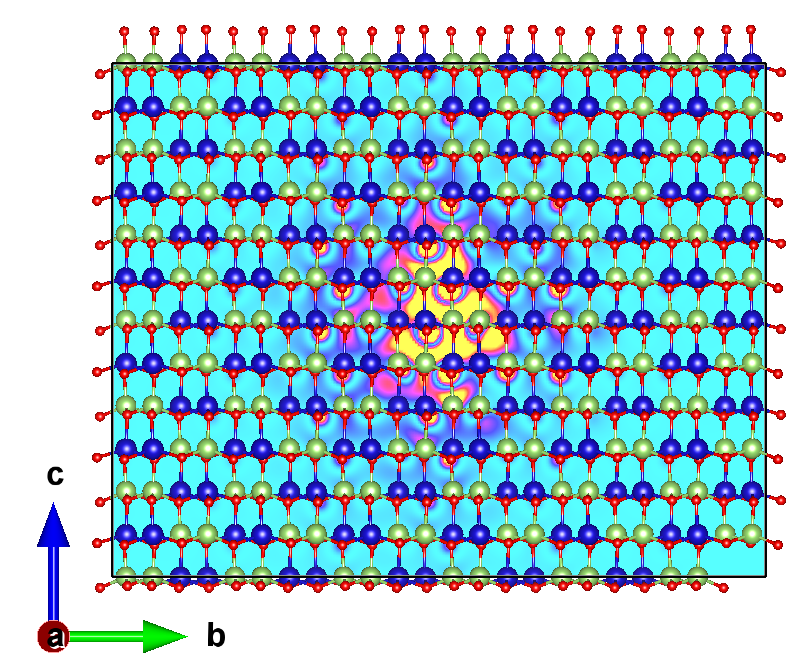}(b2)\includegraphics[width=4cm]{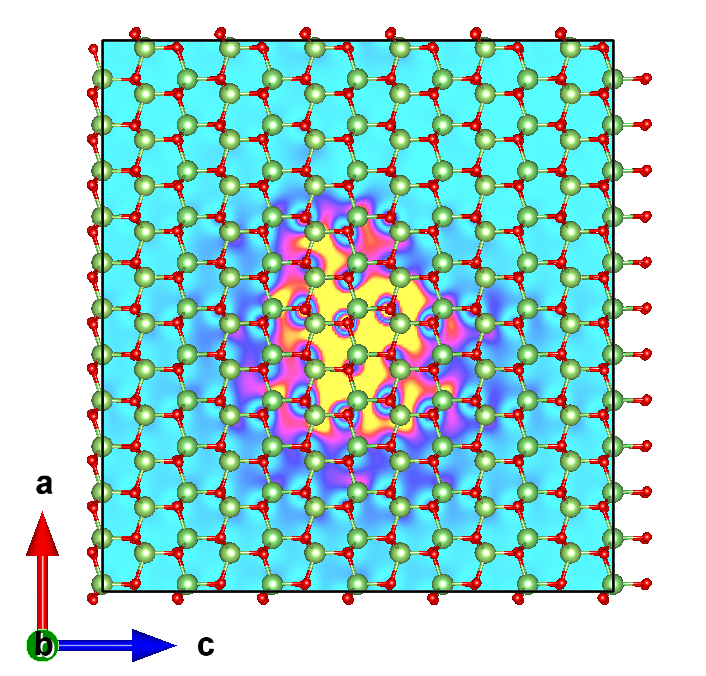}(b3)\includegraphics[width=4cm]{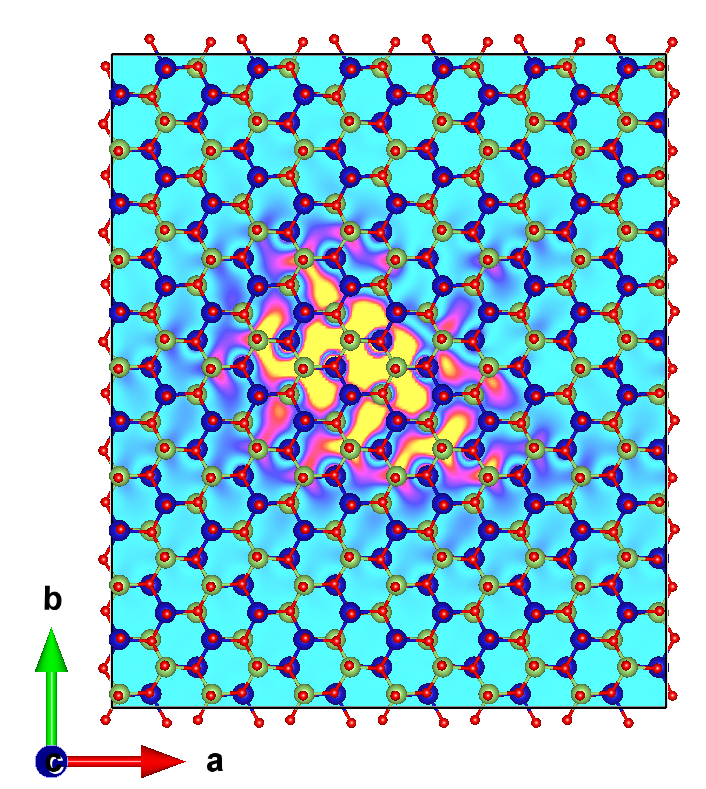} 
  (c1)\includegraphics[width=4cm]{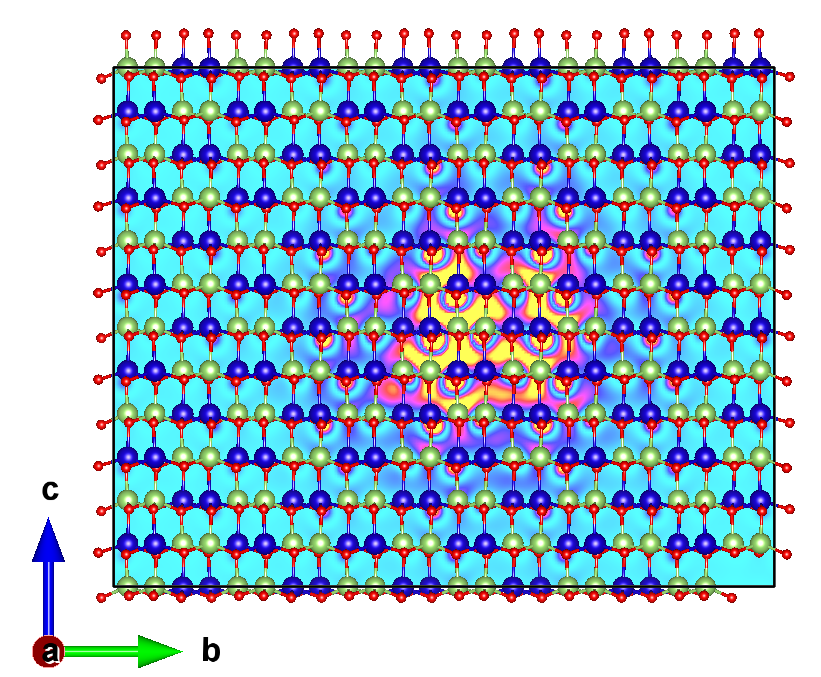}(c2)\includegraphics[width=4cm]{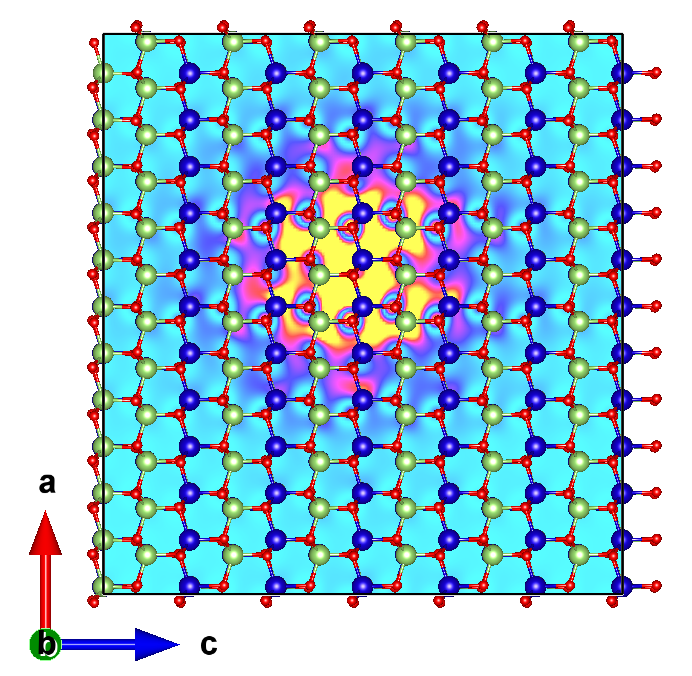}(c3)\includegraphics[width=4cm]{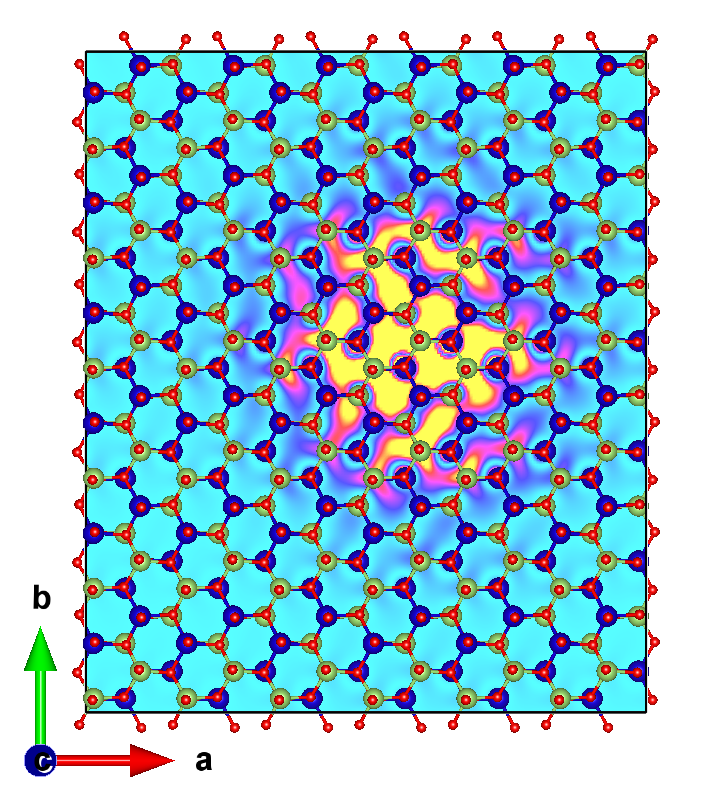}
  (d1)\includegraphics[width=4cm]{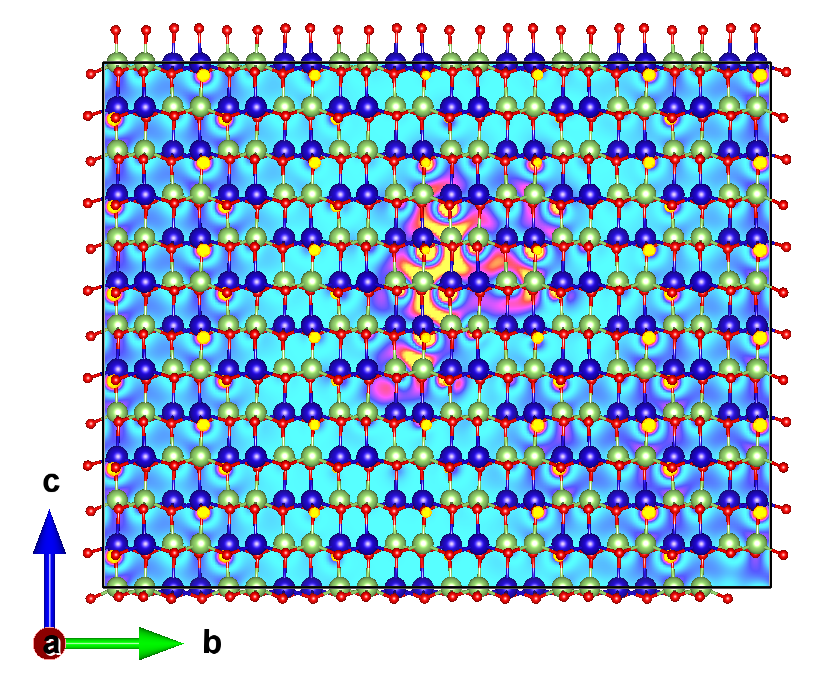}(d2)\includegraphics[width=4cm]{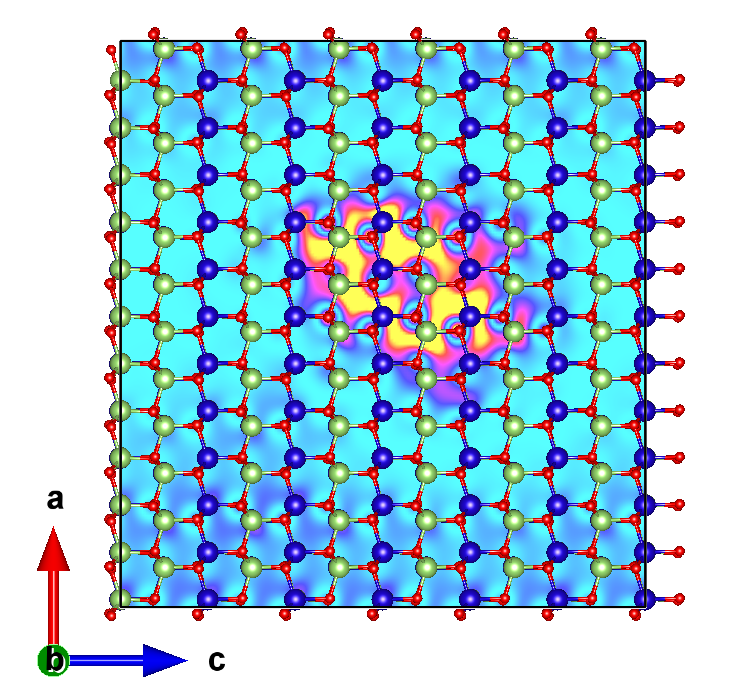}(d3)\includegraphics[width=4cm]{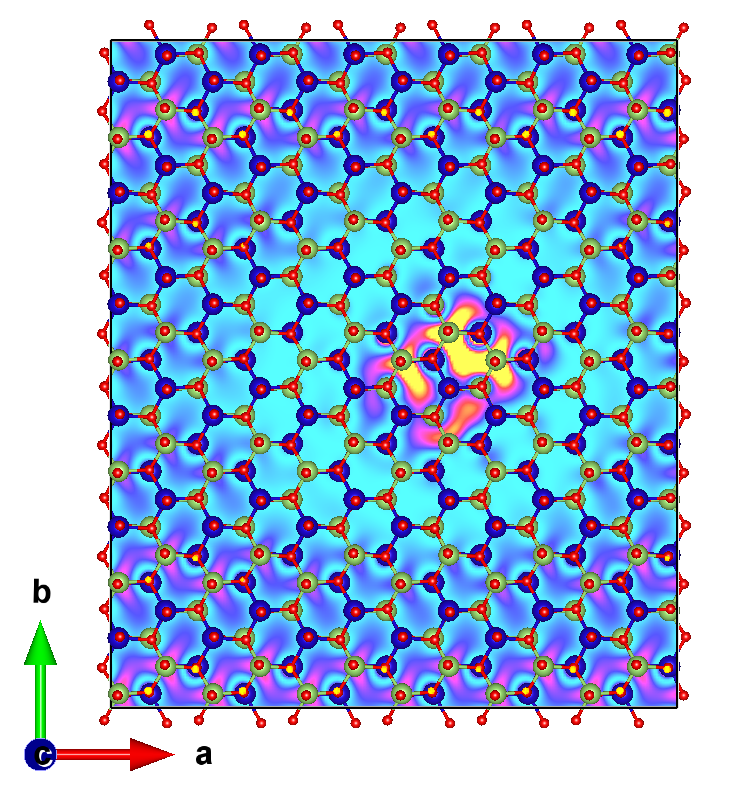}
  (e1)\includegraphics[width=4cm]{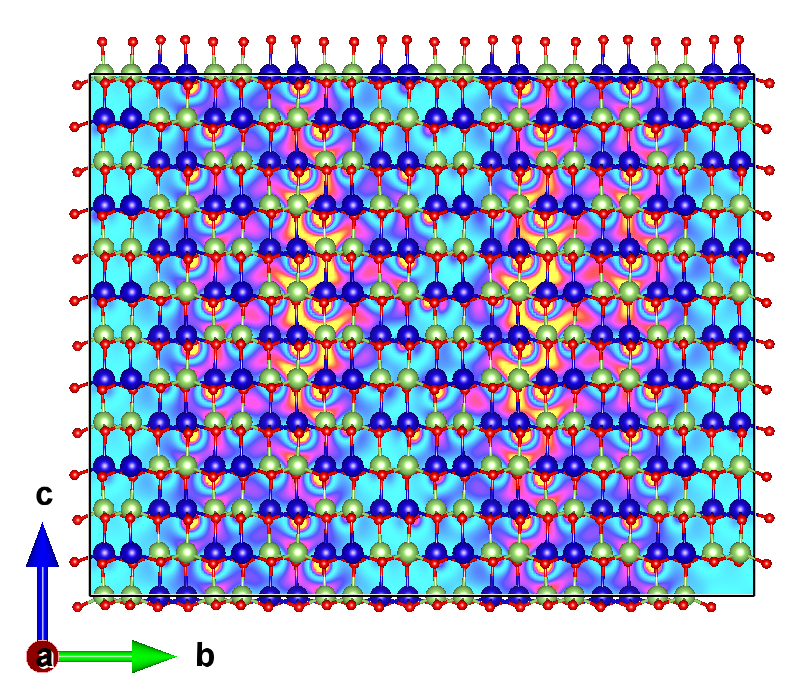}(e2)\includegraphics[width=4cm]{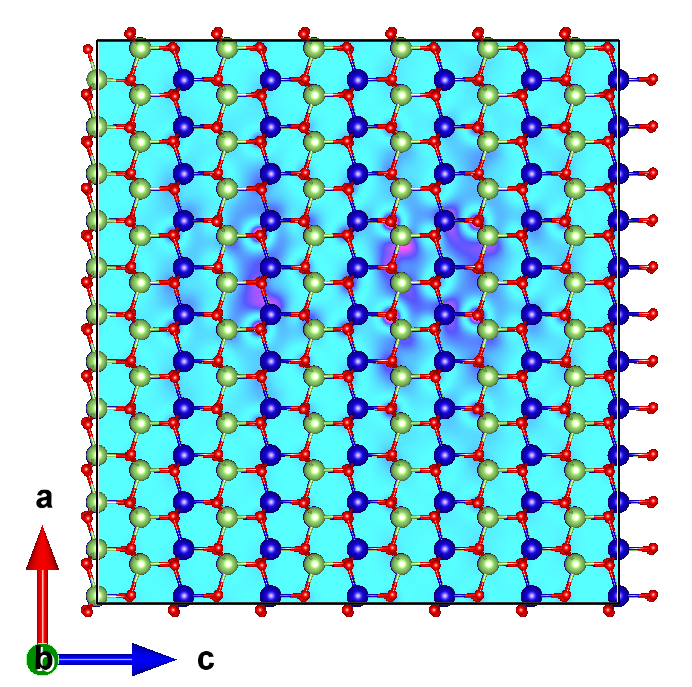}(e3)\includegraphics[width=4cm]{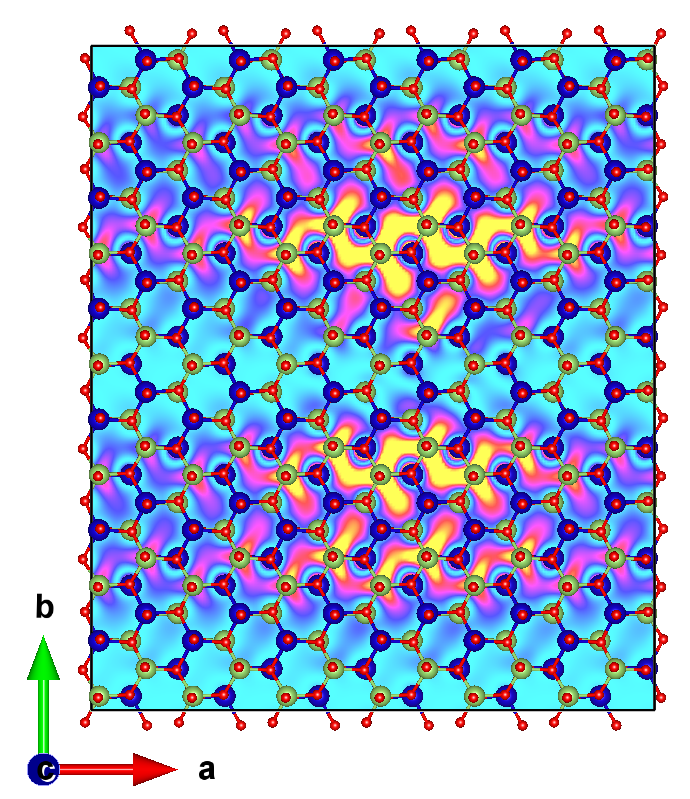}
  \caption{Sections through the center of the excitons 1--4 and 6 (of Table \ref{tabexseries}) from top to bottom and in different planes from left to right.
    The data values $d$
    are converted to a color index $T$ between 0 and 1, $T=(d-S_{min})/(S_{max}-S_{min})$ with $S_{min}=0$ and $S_{max}\approx 0.02d_{max}$. For $d>S_{max}$ the
    highest color level $T=1$ is used. 
    \label{figrealexsec}}
\end{figure*}

As a second approach, the spatial extent of the excitons in real space is illustrated
in Figs. \ref{figrealex} and \ref{figrealexsec}. Here we present isosurface plots of
$|\Psi^\lambda({\bf r}_e,{\bf r}_h)|^2=|\sum_{vc{\bf k}} A_{vc{\bf k}}^\lambda \psi_{v{\bf k}}({\bf r}_h)\psi_{c{\bf k}}({\bf r}_e)|^2$ for the hole position ${\bf r}_h$ chosen on one of the O located above a Li atom as function of the electron
position ${\bf r}_e$. In the expansion of the one-electron eigenstates,
only the smooth part of the muffin-tin orbitals represented on a
real-space mesh in the unit cell is included. The calculation
used a $6\times6\times6$ {\bf k}-point mesh and hence obtains the excitons
in a $6\times6\times6$ supercell. We plot these using the {\sc VESTA}
plotting software \cite{VESTA} and use an isosurface value, approximately
10 \% of the maximum of the function. The latter varies somewhat arbitrarily
because the smooth parts of the basis functions are not normalized.
To extract these real-space probabilities and the band weights
we reduced the number of bands
included in  the BSE calculation to only 6 valence bands
and one conduction band.  This slightly modifies the
exciton binding energies and even  how many separate excitons we obtain
as eigenvalues but we can still identify the excitons with those in
Table \ref{tabexseries}.

Because of the 3D structure, it is impractical to superpose the structure
on the isosurface plots and maintain a 3D perspective view as we do in Fig. \ref{figrealex}.
The box corresponds to a $6\times6\times6$ supercell and the hole is placed near the center.
The isosurface then shows the probability distribution of finding the electron  at a level 10 \% of the maximum.
These figures give an idea of the overall spread of the exciton and show a nonmonotonic structure in some cases.
The relation of the structure to the isosurface can be better seen in Fig. \ref{figrealexsec}
where we show sections in the a,b,c planes passing through the center of the distribution.
Here we can still see the overall spread but in addition we can see that the probability to find the electron is
larger near a few of the Ga atoms close to the hole located on an O above Li in the center of the box.
One can see that the excitons 1-3 have similar spatial extent, which
is consistent with them being $n=1$ excitons corresponding to
different valence bands. The isosurface value is chosen so as to
show sizable contributions near the atoms. The overall size is somewhat
arbitrary but clearly the excitons extend over at least 10 \AA.
This is consistent with an effective Bohr radius of $\hbar^2\varepsilon/\mu e^2$
with a reduced mass of about 0.2 and dielectric constant of about 4.

For exciton No. 4 in Table \ref{tabexseries}, which we claim is an $n=2$ bright exciton related to the
top valence band, we chose the isosurface value a bit smaller to show more
clearly that it extends farther in space. One can see that it has a central
region similar to exciton 1, then a shell of reduced intensity (corresponding
to a radial node), and  then a more extended tail where the
contributions on each atom are significantly smaller. This is more clearly seen in Fig. \ref{figrealexsec}.
This is what is expected
of a $2s$-like envelope function. The fact that the tail extends all the way
to the edges of the $6\times6\times6$ cell may indicate that this cell does not
fully capture the real-space extent of the exciton and would require a finer
{\bf k} mesh for accurate convergence.

Exciton No. 5, which is the first dark exciton in this calculation, which
uses a larger number of {\bf k} points but fewer bands,
and can be identified with exciton No. 6 in Table III,
shows two distinct  regions
with a nodal plane perpendicular to the {\bf b} axis in between.
Although we here plot only
the wave function modulo squared, giving the probability density of
finding the electron at a certain position from the chosen hole position
in the center of the supercell, we may expect this to be an odd function
as we will explicitly show below in the {\bf k}-space plots of the
real part of the $A_{vc{\bf k}}^\lambda$. This is also clearly seen in Fig. \ref{figrealexsec},
where in the {\bf b} plane through the center, the values are very small.
Some of the other excitons become more difficult to interpret
and are also deemed less well converged in terms of {\bf k} mesh or number of bands
involved, which increases as the exciton binding energies becomes smaller.
They are thus not shown here.

Finally, our third approach to analyze the excitons is to
look directly at the $A^\lambda_{vc{\bf k}}$ coefficients on a {\bf k} mesh.
We here use an $18\times18\times18$ {\bf k} mesh
but only 1 conduction band and 6 valence bands and  interpolate the results
to an  even  finer mesh. We can either inspect individual $vc$ pairs
or sum over all $vc$ pairs for a given exciton $F^\lambda({\bf k})=\sum_{vc} A^\lambda_{vc{\bf k}}$ and then display this
as function of {\bf k}.   First, we should note
that exciton energies calculated in this way is different from that in Table \ref{tabexseries}
but is deemed to be better converged in {\bf k}.
Since here we  wish to focus on the low-lying excitons,
we think it is more important
for convergence to make the ${\bf k}$ mesh as fine as possible at the expense of
including only a few bands. Hence, we focus only on the six
lowest energy excitons in the present discussion.
As discussed above,  in most cases only one pair contributes
significantly near $\Gamma$. However, as we move away from $\Gamma$ the band
plots indicate that a different band number may contribute. We will see that this leads to somewhat
intricate fine structure of the exciton eigenstates in {\bf k} space.
We here examine not only the absolute value but also the real and imaginary
parts of these envelope functions in {\bf k} space to evaluate their symmetry
by looking for sign changes.  The real and imaginary parts depend somewhat
on an arbitrary phase. So, we divide the $A^\lambda_{vc{\bf k}}$ by a constant
phase such that at the $\max_{vc{\bf k}} |A^\lambda_{vc{\bf k}}|^2$,
the $A^\lambda_{vc{\bf k}}$ becomes purely  real.

We note that the $F^\lambda({\bf k})$ functions
provide directly the 3D Fourier transform of the real-space exciton
envelope function in a Wannier exciton model. For example for a pure spherical harmonic envelope function they would
preserve the spherical harmonic character but have a radial extent in {\bf k} space given by the spherical Hankel function transform proportional to
$\tilde f_l(k)=\int_0^\infty f_l(r)j_l(kr) r^2dr$ for a radial function $f_l(r)$.
Similar plots of Wannier function envelope functions in {\bf k} space
were given for 2D MoS$_2$ by Qiu \etal\cite{Qiu16}, which
are in-plane isotropic.
However, as mentioned earlier, in the present 3D material, we
do not have a pure spherical harmonic
envelope function because of the anisotropy of
the screened Coulomb interaction and  valence band effective masses and
possibly the mixture of bands at ${\bf k}$ away from $\Gamma$.

\begin{figure*}
  \makebox[\textwidth][c]{\includegraphics[trim={0 8cm 0 9cm},clip,width=1.1\textwidth]{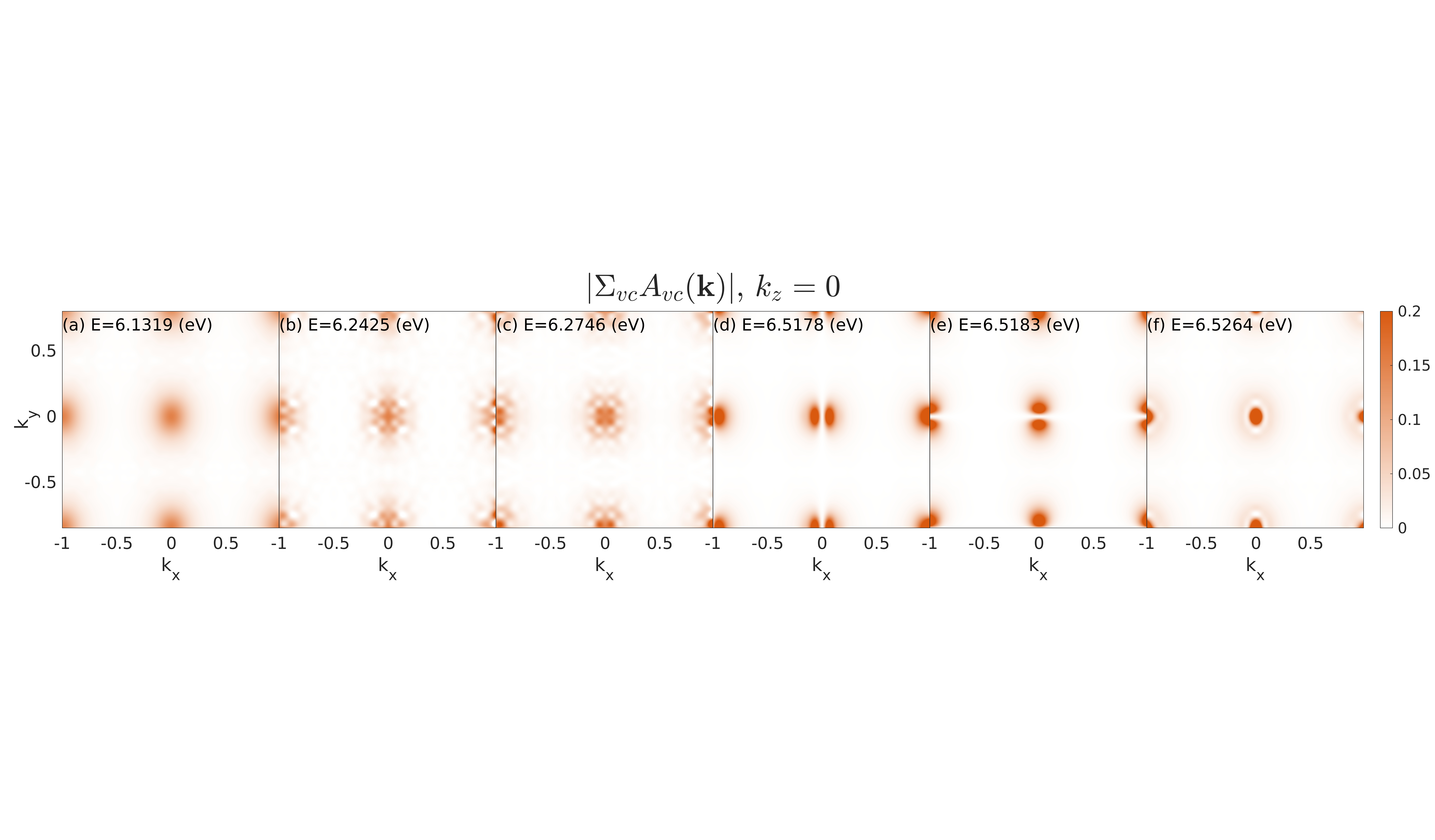}}
  \caption{Absolute value of envelope function $F^\lambda({\bf k})$ for lowest six excitons calculated with
    an $18\times18\times18$
    {\bf k} mesh. The $k_x,k_y$ are in units $2\pi/a$ with $a$ the lattice constant. \label{figkplot1}}
\end{figure*}

\begin{figure*}
  \makebox[\textwidth][c]{\includegraphics[trim={0 8cm 0 9cm},clip,width=1.1\textwidth]{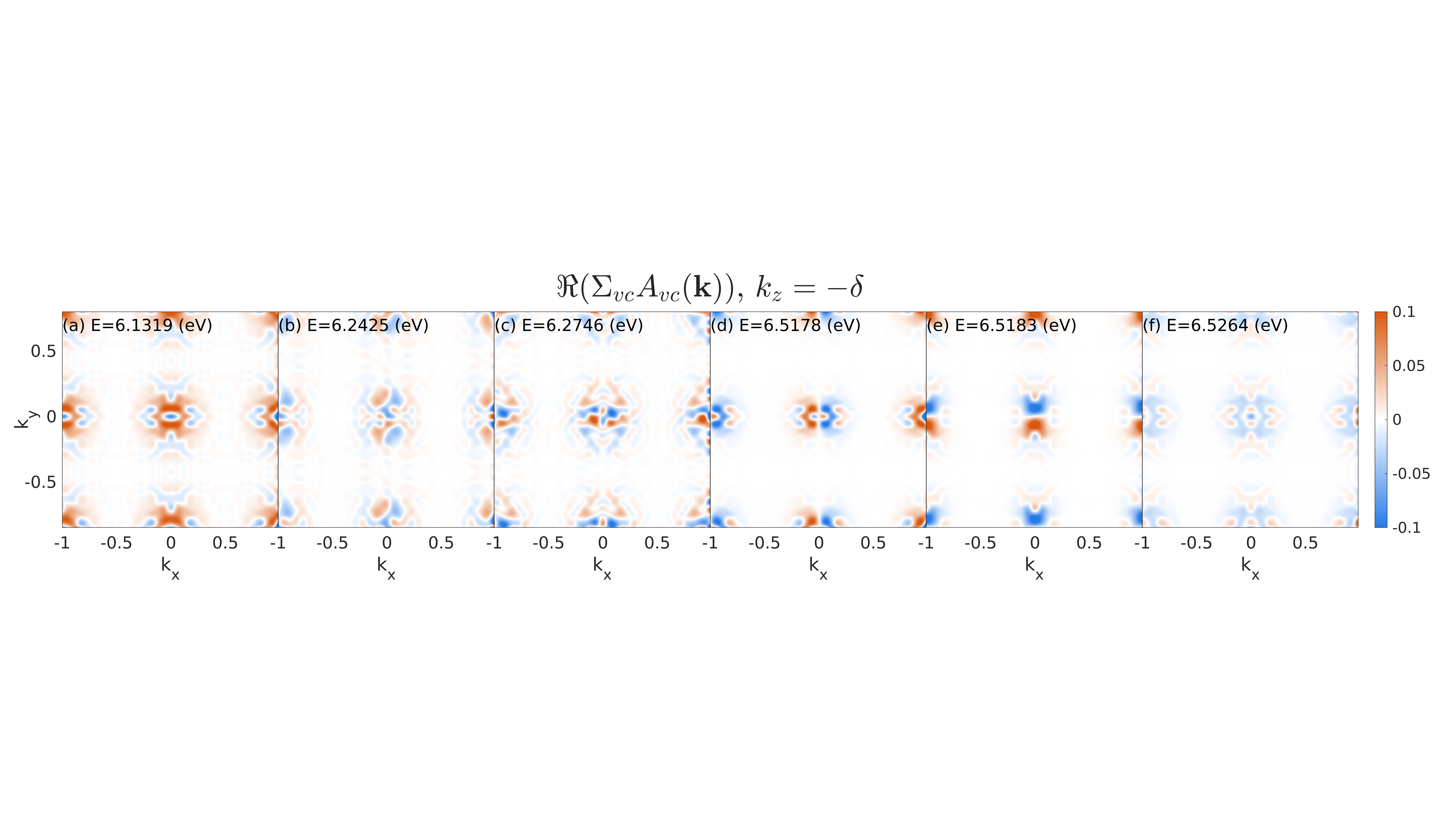}}
  \makebox[\textwidth][c]{\includegraphics[trim={0 8cm 0 9cm},clip,width=1.1\textwidth]{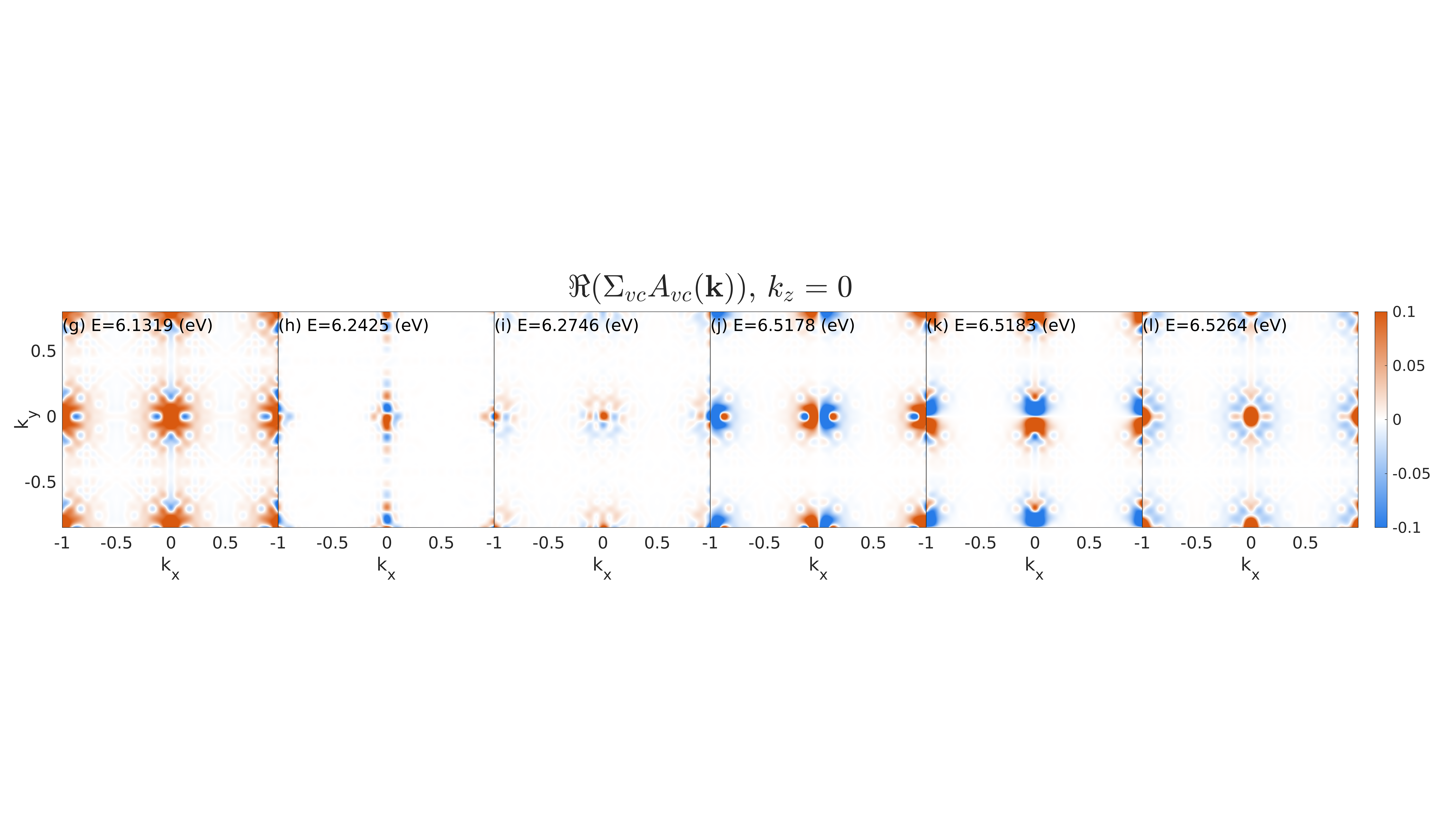}}
  \makebox[\textwidth][c]{\includegraphics[trim={0 8cm 0 9cm},clip,width=1.1\textwidth]{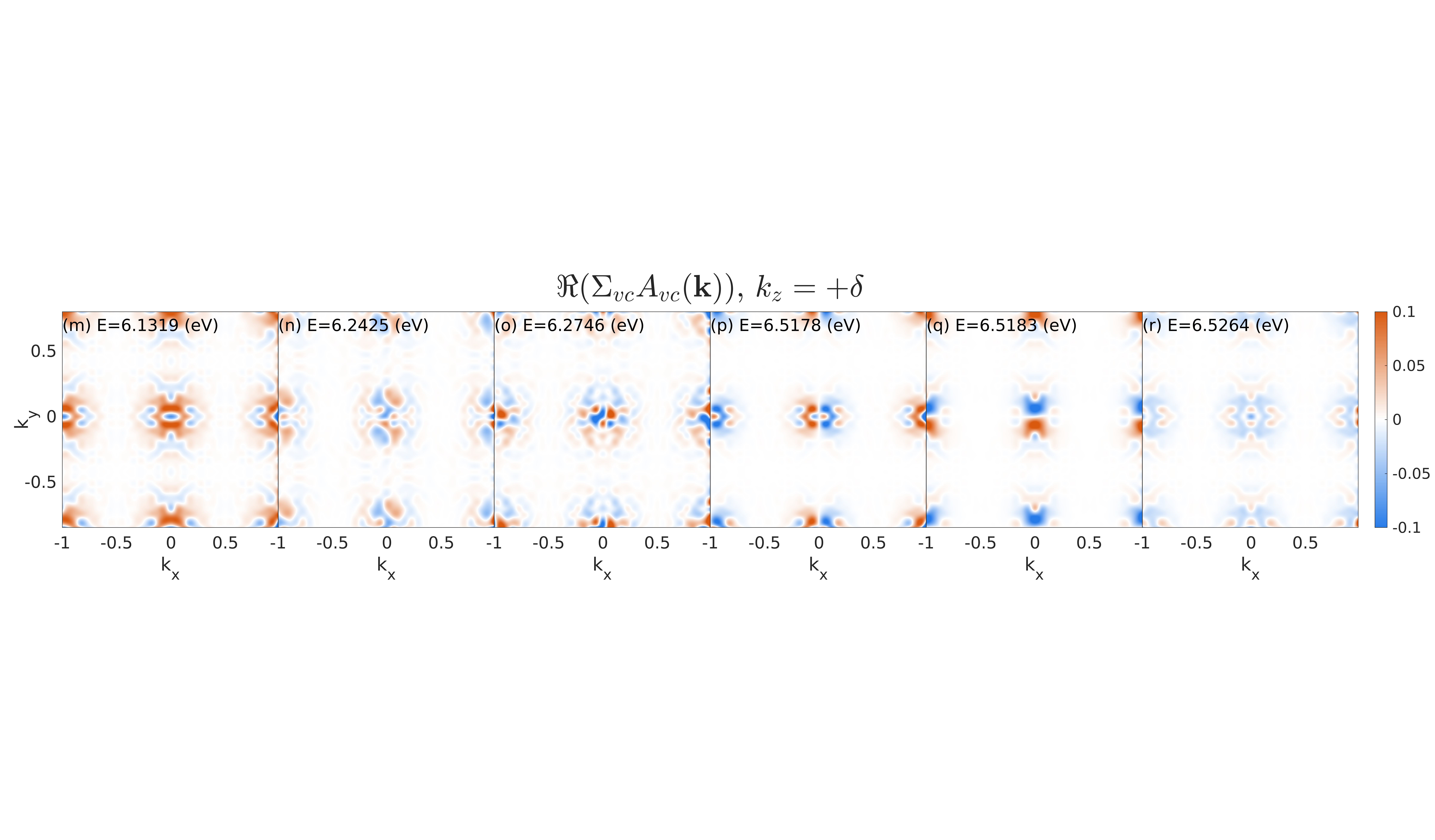}}
  \caption{Real part of the $F^\lambda({\bf k})$ in three $k_z$ planes for lowest six excitons, $k_x,k_y,k_z$ in units of $2\pi/a$. \label{figkplot2}}
\end{figure*}

Fig. \ref{figkplot1} shows the absolute value of the six lowest excitons
in the $k_z=0$ plane indicating also their eigenvalue.
We can see in Fig. \ref{figkplot1} that the first three excitons
(polarized along {\bf c}, {\bf a}, {\bf b}, respectively) have an envelope function with
similar extent in {\bf k} space and they show no radial nodes.
In other words, they are consistent with a monotonic $1s$-like function.
Nonetheless, the
second and third excitons are already seen to have a more intricate
fine structure, which will be discussed below.
Excitons 4 and 5 in this calculation turn out to be dark, and  absolute value plots show they have a node in the
$yz$ and $xz$ planes, respectively. Exciton 6 is again seen to be
even but shows a radial nodal structure, which  is the characteristic feature  of a $2s$-like exciton.
The smaller inner region in {\bf k} space indicates a larger extent in real space.
The lower intensity  second ring is of similar size as the envelope of the first exciton and results
from the orthogonality to the lower exciton envelope function. This exciton 6 here  corresponds to exciton 4 in Table III, so apparently the dark
excitons 4 and 5, which are in some sense $2p$-like, have actually lower energy than the $2s$ exciton when using a finer {\bf k} mesh, but we can see that their energies differ by less than 0.01 eV with the two $2p$ dark excitons differing by less than 0.001 eV.
It is thus not surprising that the order and number of excitons we obtain are quite sensitive to the {\bf k} mesh.

The real parts are shown in Fig. \ref{figkplot2}
for all six excitons in the $k_z=0$ and $k_z=\pm\delta k$ planes with $\delta k$ the mesh spacing in the $z$ direction.
These figures clearly show that the dark excitons 4 and 5 are odd with
respect to the mirror planes mentioned earlier, which explains why
they are dark.
A more intricate pattern of symmetries is seen in all excitons.
For example for exciton 2, we can see that the function is even under a $C_{2y}$ operation,
a two fold rotation about the $y$ axis, which changes both $k_z\rightarrow-k_z$ and $k_x\rightarrow-k_x$. It is also odd under a twofold rotation about the $z$ axis.
Although $C_{2y}$ is not a symmetry of the crystal structure, it is a symmetry of the point group of the lattice vectors, which is $D_{2h}$ and hence of {\bf k} space. We can see that in the $k_z=0$ plane it is stretched in the $y$ direction.
This is consistent with Fig. \ref{figex1k}(c) and results from the hole mass being larger in the $y$ than the $x$
direction for the second band state, which has $b_1$ (or $x$-like) symmetry.  This means the function will be spread
out more in the $x$ direction in real space, as can also be seen in Fig. \ref{figrealex}(c).
From Fig. \ref{figex1k} one can see that this exciton will also have contributions from the first and third band
beyond the band crossings in {\bf k} space.  This may account for the complex superposition of different patterns.
The same is true for each of these excitons. Nonetheless, one can see that the first and sixth excitons are fully $a_1$ symmetric. Their patterns also look quite similar at larger $k$ but differ closer to $\Gamma$. Clearly, the Wannier exciton
model based on spherical symmetry does not quite hold because of the more complex mixing of Bloch states of
different {\bf k} in the BSE theory and the anisotropies of the present system but still provides an approximate guidance
to understand these excitons.

\section{Conclusions}
The first conclusion of this work is that the quasiparticle bandgap of
$\beta$-LiGaO$_2$ calculated previously  in \cite{Radha21}
has to be revised for three reasons. First, better self-consistency
convergence of the QS$GW$ gap  increases the gap from 6.46 eV to 7.22 eV.
Second, adding ladder corrections to the polarizability leads to
a QS$G\hat W$ gap of 7.02 eV, where the self-energy is reduced by about 5 \%
rather than the canonical 20 \%.
Third, the electron-phonon coupling band gap renormalization estimated
there was meanwhile fully calculated in \cite{Dangqi22} and
gives a larger correction of -0.36 eV. Considering all these, the
quasiparticle gap becomes  6.66 eV.
However, exciton binding energies are found to be about 0.7 eV
for the ground-state excitons related to each valence band maximum and
the conduction band minimum. Taken together, this places
the optical exciton gap at
5.96 eV with a dipole-allowed transition with polarization along the ${\bf c}$ axis, followed by a 6.06 eV exciton polarized along {\bf a} and 6.12 eV
along {\bf b}. These results are in excellent agreement with
recent spectroscopic ellipsometry\cite{Tumenas17} and photoluminescence
excitation results \cite{Trinkler17,Trinkler22}.
These results were obtained by extrapolating the calculated
exciton energies as function of the inverse of the number of {\bf k} points
in the Brillouin zone sampling to zero.
Since only electronic screening is included in the BSE
calculations done here, this excellent agreement with experiment
suggests that indeed only electronic (as opposed to lattice) screening  affects the exciton binding energy.
The static real (electronic-only) dielectric constant $\varepsilon_1(\omega=0)$ is also
found to be in good agreement with experiment, suggesting that the method captures the correct
amount of screening rather well by including the ladder diagrams, and that these low-order diagrams are sufficient to well capture both the one-particle Green's function and the two-particle dielectric function.  
The excellent agreement could be an artifact of error cancellation in the various approximations made, in particular the
  use of static, RPA $W$ in the vertex for the BSE, use of the TDA, and the omission of higher order diagrams.  We have
  made a few checks of the TDA and RPA $W$ in several weakly correlated systems, and found the effects to be relatively
  small although not completely negligible.  That being said, the high degree of fidelity in one- and two-particle properties for many kinds of materials and the
  consistency between one- and two-particle properties suggest that if results are improved by error cancellation, it is not entirely
  fortuitous but occurs for some reason akin to the $Z$-factor cancellation in the self-energy noted earlier.

Further examination of the excitons below the gap
reveals that the BSE gives approximately a  Rydberg-like series of excitons
associated with each band edge. However, it deviates from the usual
Rydberg series because of the anisotropy of the band states involved in
the exciton and the Coulomb energy  and the band mixing in the BSE.
Several dark excitons were also found
and they were shown to be associated with the same bands but with a
nodal plane and therefore not fully symmetric envelope function
within the point group of the system, which explains
why they become dark.  Rather intricate patterns of the exciton envelope functions
were revealed by using a sufficiently fine {\bf k} mesh.

\acknowledgements{The work at CWRU was supported by the U.S. Department of
  Energy Basic Energy Sciences (DOE-BES) under grant No. DE-SC0008933.
  Calculations made use of the High Performance Computing Resource in the Core Facility for Advanced Research Computing at Case Western Reserve University and the Ohio Supercomputer Center.
  D.P. and M.v.S. were supported by the DOE-BES, Division of Chemical Sciences, under Contract No. DE- AC36-08GO28308.
}

 \bibliography{ligao2,dft,gw,lmto}
\end{document}